\documentclass[fleqn,usenatbib]{mnras}
\usepackage{newtxtext,newtxmath}
\usepackage[T1]{fontenc}
\usepackage{ae,aecompl}
\usepackage[normalem]{ulem}
\usepackage{graphicx}	
\usepackage{amsmath}	
\usepackage{newtxtext,newtxmath}
\usepackage{array, booktabs, makecell} 
\usepackage{comment}
\usepackage{url}

\newcommand{\MAXI}{{MAXI\, J1820+070\,}\,}

\title[Warped Accretion Disc in J1820]{Large optical modulations during 2018 outburst of MAXI~J1820+070 reveal evolution of warped accretion disc through X-ray state change}\title[Warped Accretion Disc in J1820]{Large optical modulations during 2018 outburst of MAXI~J1820+070 reveal evolution of warped accretion disc through X-ray state change}

\author[J K Thomas et al.]{Jessymol K. Thomas,$^{1}$\thanks{E-mail: jessy@saao.ac.za}, Philip A. Charles$^{2}$,
David A. H. Buckley$^{1,3}$, Marissa M. Kotze $^{1,3}$, 
\newauthor {Jean-Pierre Lasota$^{4,8}$, Stephen B. Potter$^{1,6}$, James F. Steiner$^{7}$, John A. Paice$^{2,5}$}
\\ \\
$^{1}$ South African Astronomical Observatory, Observatory Road, Observatory, 7925, Cape Town, RSA\\
$^{2}$ Department of Physics $\&$ Astronomy, University of Southampton, Southampton SO17 1BJ, UK\\
$^{3}$Department of Physics, University of the Free State, PO Box 339, Bloemfontein 9300, South Africa\\
$^{4}$ Institut d’Astrophysique de Paris, CNRS et Sorbonne Université, UMR 7095, 98bis Bd Arago, 75014 Paris, France\\
$^{5}$ Inter-University Centre for Astronomy and Astrophysics, Pune, Maharashtra 411007, India\\
$^{6}$ Department of Physics, University of Johannesburg, PO Box 524, Auckland Park 2006, South Africa\\ 
$^{7}$ Center for Astrophysics, Harvard-Smithsonian, 60 Garden St. Cambridge, MA 02138, USA\\
$^{8}$ Nicolaus Copernicus Astronomical Center, Polish Academy of Sciences, Bartycka 18, 00-716 Warsaw, Poland\\}
 
\date{Accepted XXX. Received YYY; in original form ZZZ}

\pubyear{2020}

\begin{document}
\label{firstpage}  
\pagerange{\pageref{firstpage}--\pageref{lastpage}}
\maketitle


\begin{abstract}
The black-hole X-ray transient \MAXI (=ASSASN-18ey) discovered in March 2018 was one of the optically brightest ever seen, which has resulted in very detailed optical outburst light-curves being obtained. We combine them here with X-ray and radio light-curves to show the major geometric changes the source undergoes.  We present a detailed temporal analysis that reveals the presence of remarkably high amplitude ($>$0.5 mag) modulations, which evolve from the superhump (16.87\,h) period towards the presumed orbital (16.45\,h) period.  These modulations appear $\sim$87d after the outburst began, and follow the Swift/BAT hard X-ray light-curve, which peaks 4 days before the radio flare and jet ejection, when the source undergoes a rapid hard to soft state transition. The optical modulation then moves closer to the orbital period, with a light curve peak that drifts slowly in orbital phase from $\sim$0.8 to $\sim$0.3 during the soft state. We propose that the unprecedentedly large amplitude modulation requires a warp in the disc in order to provide a large enough radiating area, and for the warp to be  irradiation-driven.  Its sudden turn-on implies a change in the inner disc geometry that raises the hard X-ray emitting component to a height where it can illuminate the warped outer disc regions.
\end{abstract}

\begin{keywords}
astronomical data bases: miscellaneous --- accretion --- accretion discs ---  stars: individual (MAXI\,J1820+070) --- X-rays: binaries
\end{keywords}


\section{Introduction}\label{sec:1}

The X-ray transient, MAXI J1820+070, was discovered via its X-ray outburst in March 2018 \citep{ATel11399}, although it was also recorded as an optical transient, ASASSN-18ey, a few days earlier \citep{Denisenko18}. It became one of the brightest (in both X-rays and optical) of these objects ever seen \citep{Atel11488,Shidatsu18,ATel11421}, and remained active and bright for many months, undergoing X-ray state transitions and several subsequent outbursts.  Also found to be a powerful, steady radio source, \citet{Bright2020} detected a large radio flare and jet ejection at the time of the first hard-to-soft state transition, following which the radio flux was substantially reduced.

Its large increase in brightness (from a quiescent V$\sim$19 to $\leq12$; \citealt{Russell2019AN,ATel12534}) immediately indicated that it was a Low-Mass X-ray Binary (LMXB) in which a cool, evolved low-mass donor transfers material by Roche-lobe overflow via an accretion disc onto a compact object.  Based on its X-ray properties \citep{Shidatsu18} it was suspected to be a black-hole, and the source has been extensively observed throughout the outburst, from both ground- and space-based facilities.  The rare ($\sim$decades apart) outbursts only occur when the disc becomes sufficiently massive to undergo a transition to a hot, viscous state, which then results in a higher mass-transfer rate onto the BH. These black-hole X-ray binaries (BHXBs) provide superb laboratories within our Galaxy for studies of the physics of BH environments and the behaviour of X-ray irradiated accretion discs.  For recent reviews of BHXBs and their properties see e.g. \cite{Casares17}.

The unusually high visual brightness of MAXI J1820+070 (hereafter J1820), only exceeded by the prototypical BHXBs A0620-00 and V404 Cyg, combined with its appearance at the start of the observing season and its low extinction (it is at Galactic latitude +10${^{\circ}}$), has allowed for the generation of a remarkably detailed outburst light-curve by the AAVSO.  Some of these data were reported by \cite{Patterson2018ATel11756}, who detected a large amplitude ($\sim$0.5 mag) 16.87\,hr modulation, beginning approximately 75d after the start of outburst (with no such variation present earlier). They considered this to be indicative of the orbital period, although conceded that it might also be a ``superhump'', a feature originally found in the SU UMa dwarf nova sub-class of cataclysmic variables (CVs), which is always a few percent longer than $P_{\rm orb}$ and is commonly interpreted as being due to a precessing accretion disc (see \citealt{WhitehurstKing1991}).

What is remarkable about J1820 is the extremely high amplitude of optical modulation seen by \cite{Patterson2018ATel11756} at the superhump period of 16.87\,hr.  Such modulations have been seen before in BHXBs \citep{O'DonoghueCharles1996}, and explained by \cite{Haswell2001} as arising due to X-ray irradiation of a disc whose area is changing on the precession period. However, those superhumps were of much lower amplitude ($\leq$ 0.2 mag).

Following its transition to quiescence in 2019 (although there have been subsequent shorter outbursts), \cite{Torres19,Torres2020} obtained the optical radial velocity curve and rotational broadening of the K4/5 donor star, finding the orbital period to be 16.45\,hr and obtaining a mass function of 5.2\,M$_{\odot}$.  This confirmed that the \cite{Patterson2018ATel11756} modulation was indeed a superhump.  From these data \citep{Torres2020} deduced the binary parameters of J1820 to be a 0.6\,M$_{\odot}$ donor orbiting a 8.5\,M$_{\odot}$ BH with a high orbital inclination (in the range 66--81$^{\circ}$, with 73.5$^{\circ}$ preferred), thereby confirming its BHXB status although \citet{PAtri2020} inferred a slightly lower value (63$^{\circ}$) based on the radio jet properties.

Furthermore, at only 3.0$\pm$0.3 kpc distance \citep{PAtri2020}, there has been extensive coverage of this outburst at X-ray and radio wavelengths, producing a wealth of results (see e.g. \citealt{Stiele2020} and references therein), and \citet{Bright2020} and \citet{Shaw21} place its X-ray/radio properties into the context of the BHXB population. There has been particular interest in the X-ray spectra and timing behaviour in the interval leading up to the very rapid (only a few days) hard-to-soft state transition, $\sim$120d into the outburst. That is because the canonical model of BHXBs has a truncated inner accretion disc surrounding a very hot, hard X-ray emitting region during the hard state, followed by the disc radius decreasing towards the innermost stable circular orbit as it enters the soft state.  The details of this process are still highly controversial.  

The aim of this paper is therefore to bring together for the first time an extremely detailed AAVSO light-curve of the 2018-19 main outburst together with its X-ray and radio behaviour.  In particular, we explore the properties of the optical modulation as it evolves leading up to, and through, the state transition.

\section{Observations}\label{sec:2}

The optical observations of J1820 used in this paper were carried out by the American Association of Variable Star Observers (AAVSO), the Southern African Large Telescope (SALT) and the South African Astronomical Observatory (SAAO) 1\,m telescope. The X-ray light curves obtained by NICER, the BAT and XRT instruments aboard the Niel Gehrels $\it Swift$ Observatory (hereafter \textit{Swift}/BAT and \textit{Swift}/XRT respectively), and MAXI, together with the radio coverage by AMI, MeerKAT, eMERLIN and VLA are also discussed here to provide full multi-wavelength observations of the overall temporal evolution of J1820 for $\sim$200d following the 2018 outburst.  For convenient reference to different times within the outburst, we adopt the same day number scheme as used by \cite{Stiele2020}; i.e. day 0 is 2018 Mar 11 0h UT = MJD 58188, just before the first triggering of X-rays from J1820 \citep{ATel11399}.

\subsection{AAVSO data}\label{sec:2.1} 

The optical brightness of J1820, reaching V$\sim$12, meant that the source could be accurately and reliably monitored by the global network of the AAVSO for extended periods of time. This provided an essentially continuous light curve throughout the 2018 main outburst. The AAVSO observers providing the most accurate brightness estimates typically used CCD cameras on 0.2-0.4 m telescopes, with time resolutions of $\sim$5-60 secs, more than sufficient for the timescales of interest here.  We accessed the J1820 V-band data from the online AAVSO database \footnote{\url{https://aavso.org/aavso-international-database-aid}}, which contains the details of each observation (including comparison stars used), selecting only those data with quoted errors $<$0.02 mag, which we then used for the temporal analysis. This yielded $\sim$370,000 data points, which are plotted in Figure \ref{fig:LC}.

\subsection{SAAO/SALT}

\begin{figure*}
\includegraphics[width=\textwidth]{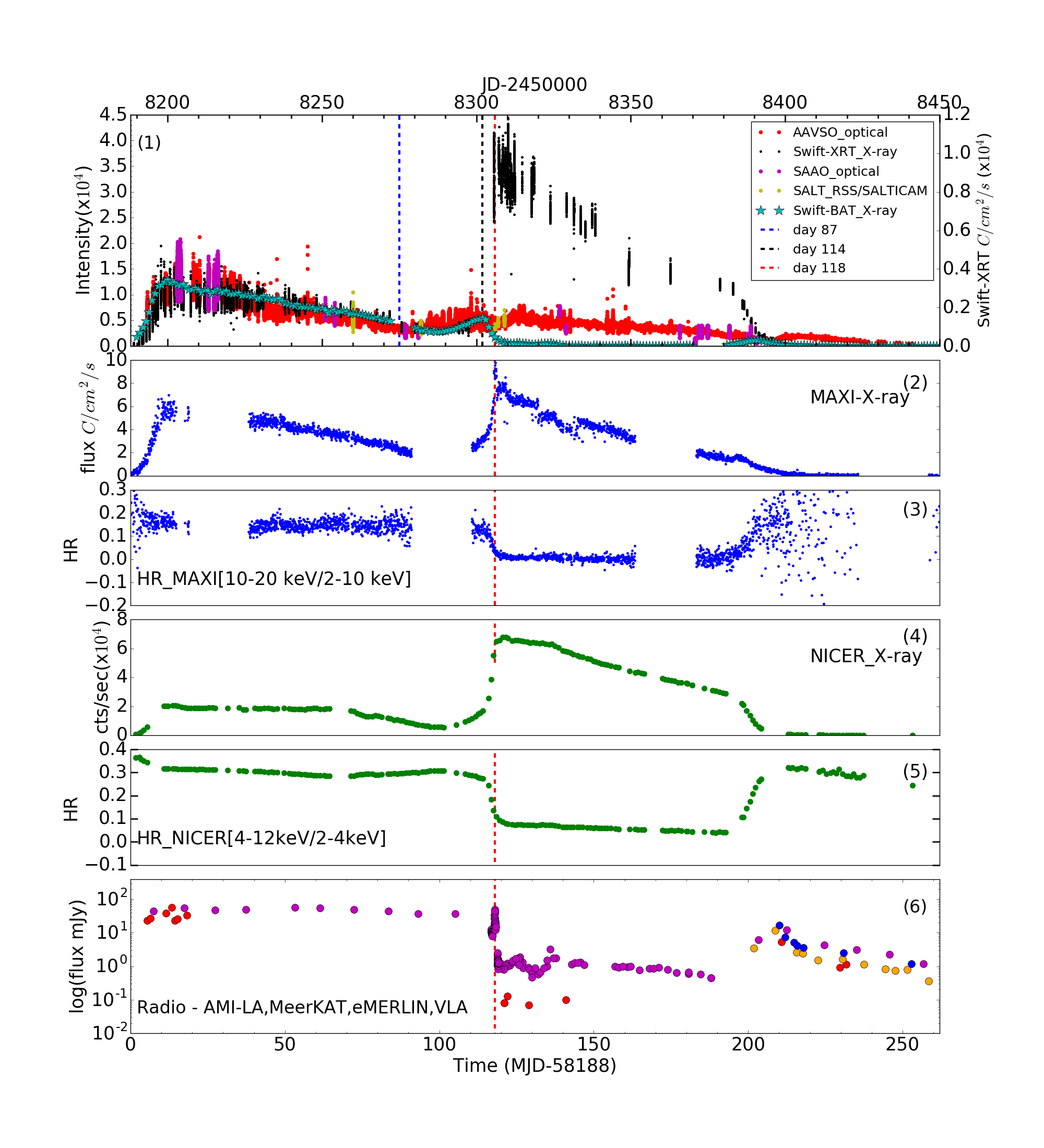}
\caption{\label{fig:LC} Multi-wavelength light-curves of J1820.  From top to bottom: {\it (1)} Optical (AAVSO - red; SAAO 1\,m - magenta; SALT - yellow) and X-ray \textit{Swift-XRT} - black; \textit{Swift-BAT} - cyan. Optical magnitudes were converted to an arbitrary intensity scale; \textit{Swift-XRT} count rates (right Y-axis) are offset by $+9000$.
 {\it (2)} MAXI count rate.  {\it (3)} MAXI Hardness Ratio (10-20\,keV/ 2-10\,keV).  {\it (4)} NICER count rate. {\it (5)} NICER Hardness Ratio (4-12\,keV/2-4\,keV).  {\it (6)} Radio fluxes from AMI-LA (magenta; 15.5\,GHz), MeerKAT (orange; 1.28\,GHz), eMERLIN (red; 1.5, 5\,GHz) and VLA (blue; C-band) are plotted in the bottom panel (see \protect\cite{Bright2020}, \protect\cite{Homan20} for details). 
Three vertical dashed lines mark key intervals during this outburst: Days 87 (blue), the beginning of large amplitude optical modulations; 114 (black), peak of the {\it Swift}/BAT secondary maximum; 118 (red), time of the AMI radio flare, jet ejection and X-ray state change from hard to soft - see Discussion section.}
\end{figure*}

We obtained high speed photometry of J1820 using the SAAO 1m telescope for 20 nights, from 25 March to 29 September 2018, equipped with the Sutherland High Speed Optical Camera (SHOC, \citealt{2013SHOC}), a frame transfer EM-CCD camera (an Andor iXon888 camera).  The light curves derived from the 200\,ms images were produced from differential aperture photometry using the SHOC data reduction pipeline\footnote{\url{https://shoc.saao.ac.za/Pipeline/SHOCpipeline.pdf}}, and these data are included in Figure \ref{fig:LC}, albeit in time-averaged form. 
 
Photometry of J1820 was also performed with the Southern African Large Telescope (SALT) on 5 days between May and July 2018. The observations on 13 June and 7 July 2018 used SALT's imaging spectrograph - the Robert Stobie Spectrograph (RSS). There is also SALTICAM, that acts as an  acquisition camera and fast science imager for SALT, and was used for the 22 May, 8 and 10 July observations. The reduced data are also plotted in Figure \ref{fig:LC}, again in time-averaged form, but after converting from magnitudes into an arbitrary intensity scale. The intrinsically high time resolution data from both SALT and SHOC will be presented in a subsequent paper.

\subsection{X-ray data: \textit{Swift}, NICER, MAXI}\label{sec:2.4} 

In addition to its continuous monitoring with the BAT (15--150\,keV), \textit{Swift} observed J1820 extensively with targeted XRT (0.3--10\,keV;  \citet{Burrows05}) observations throughout the 2018 outburst \citep{Stiele2020}, and we used the \textit{Build Swift-XRT} products (from the UK \textit{Swift} Science Data Centre) to construct an X-ray light-curve for use here.  These data are presented in Figure \ref{fig:LC}, together with X-ray observations reported by NICER (0.5--12\,keV; see \citet{Gendreau12}) from March to September 2018, and the MAXI (2--20\,keV; see \citet{Matsuoka2009}) X-ray light-curve data collected from the MAXI data archive centre. 

\section{Temporal analysis}\label{sec:3}

As already noted by \cite{Patterson2018ATel11756}, J1820 exhibited large, photometric variations with a period of $0.703\pm0.003$\,d (subsequently revised to 0.6903d in \citealt{Patterson19}), and these are readily visible in the AAVSO light-curve, but only during certain intervals.  Since the orbital period has now been accurately determined (\citealt{Torres19}; \citealt{Torres2020}) to be $P_{\rm orb} = 0.68549\pm0.00001$\,d, we undertook a Lomb-Scargle periodogram analysis of this rich AAVSO optical dataset to study in detail these variations of J1820 throughout the $\sim$1 year outburst. The relevant periodicities in which we are interested are the orbital, superhump ($P_{\rm sh}$) and precession ($P_{\rm prec}$) periods, where $P_{\rm prec}$ is the beat period between the first two.  We used a dynamical power spectrum (DPS) analysis to reveal how these periodicities of J1820 evolve with time (as used e.g. by \citet{Clarkson03} and \citet{Kotze12}).  The data coverage is sufficiently extensive to also allow for phase-folding of the light curves as a function of time during the outburst.

\subsection{Presence of orbital and superorbital modulations}

 We have used the \textquotesingle{Lomb-Scargle (LS) Periodogram}\textquotesingle{} from gatspy.periodic \footnote{Gatspy, created by Jake VanderPlas, is a collection of tools for analyzing astronomical time series data in Python \citep{VanderPlas2015}.}, to perform the period analysis of our optical and X-ray light-curves. The Lomb-Scargle periodogram (see \citet{Lomb1976} and  \citet{Scargle1982}) is a commonly used statistical tool for detecting periodic signals in unevenly spaced observations.

\begin{figure}
\includegraphics[width=0.52\textwidth]{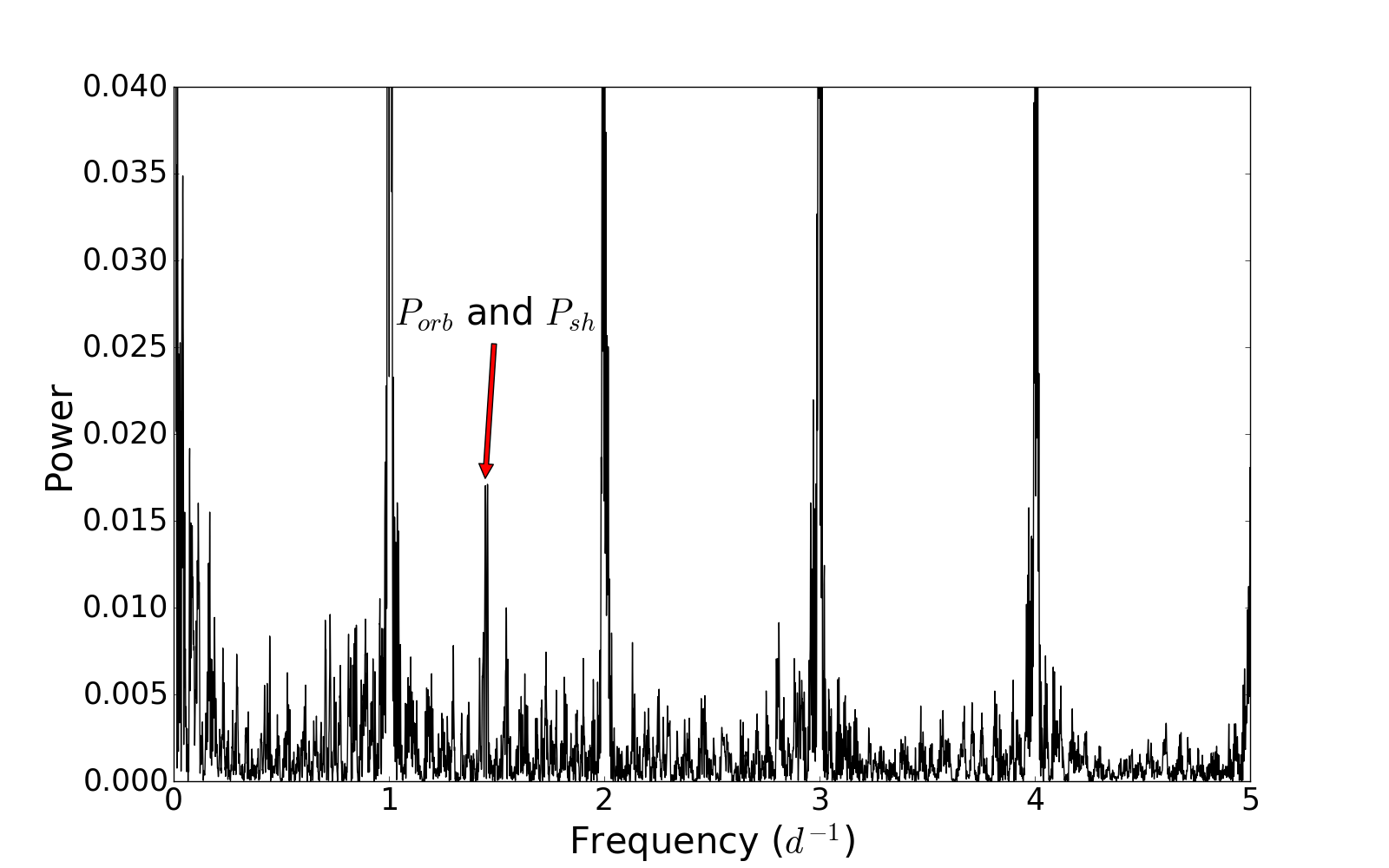}
\includegraphics[width=0.52\textwidth]{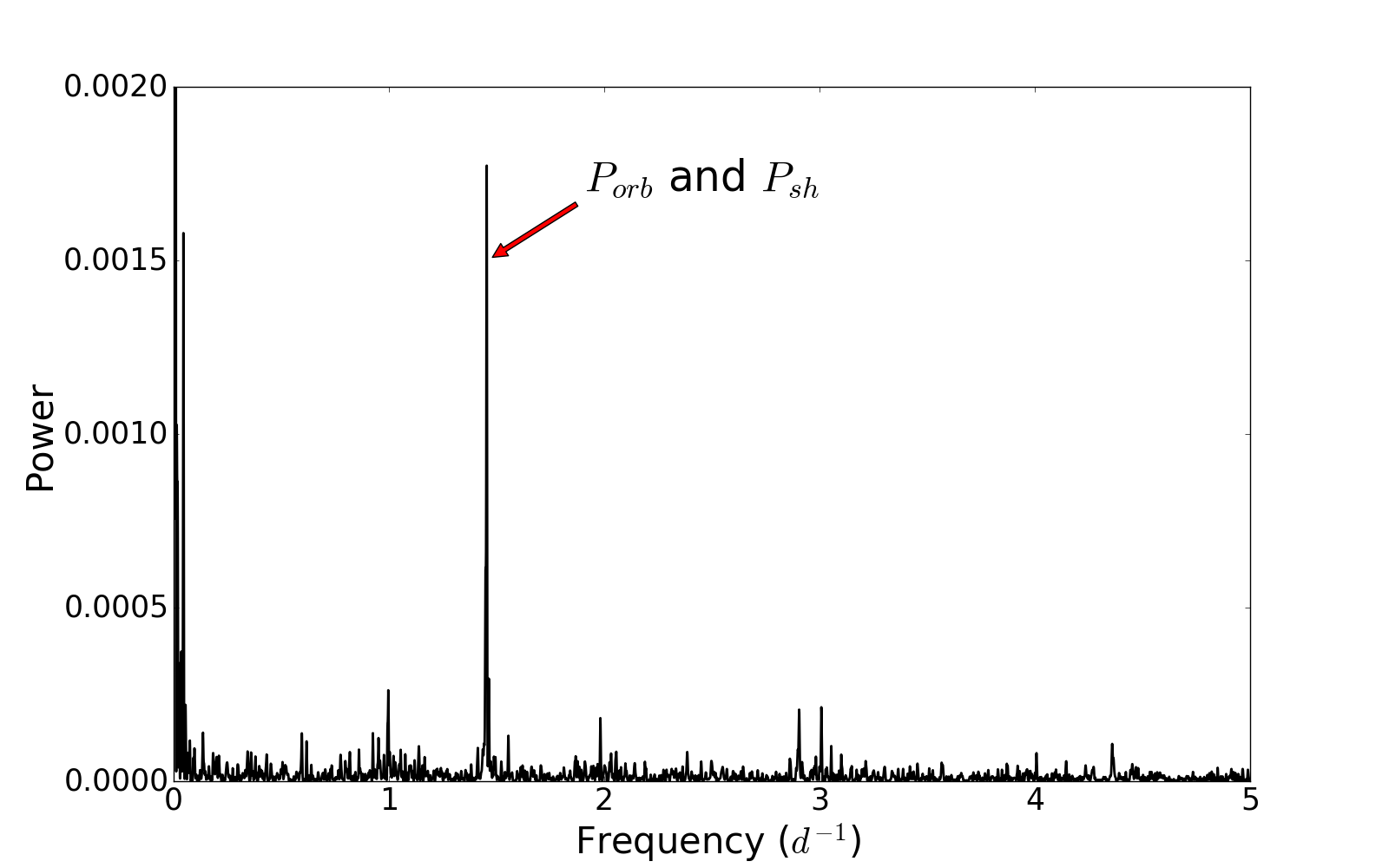}
\includegraphics[width=0.52\textwidth]{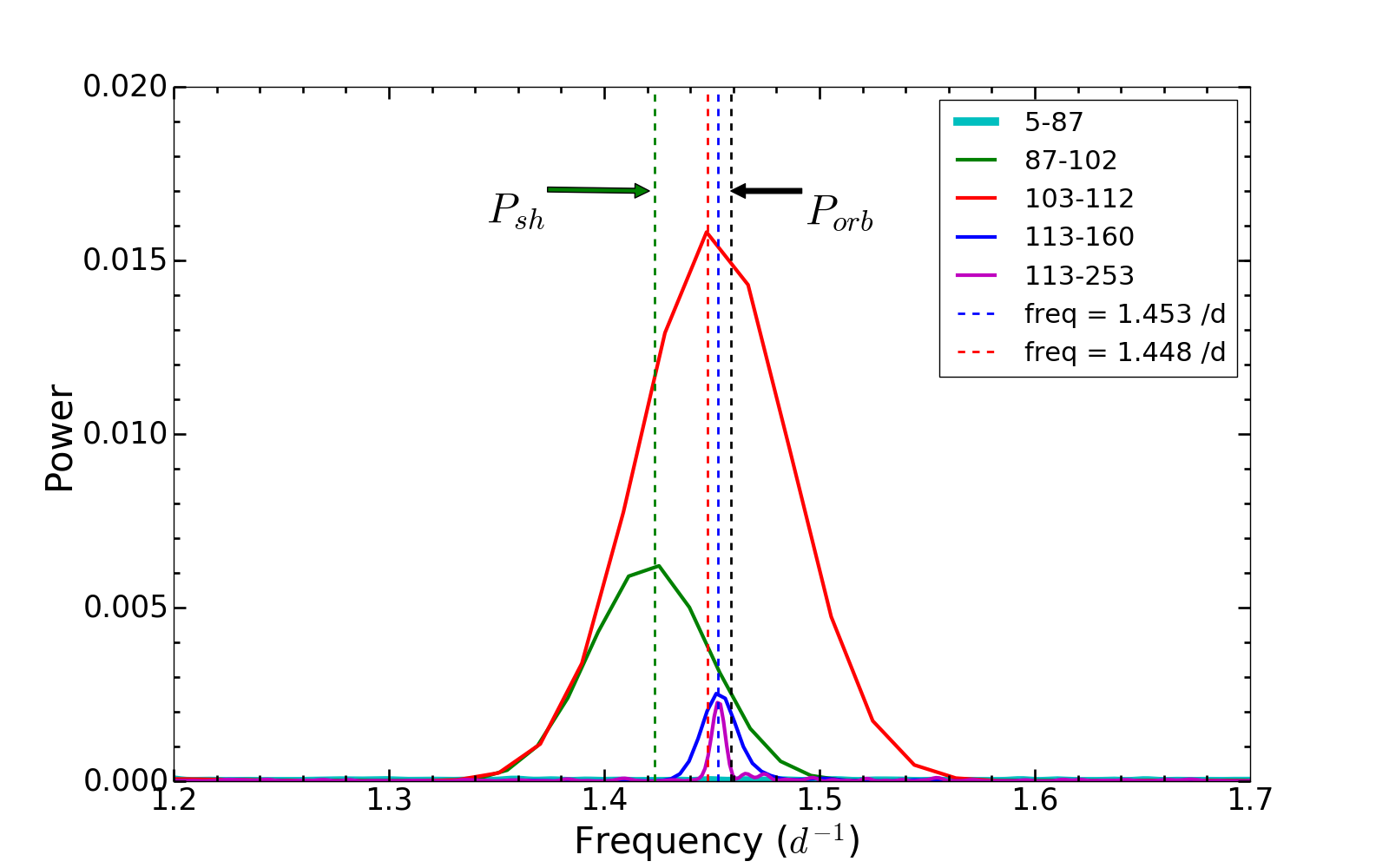}
\caption{\label{fig:LS_AAVSO} Periodograms of the AAVSO 2018 V-band light-curve of J1820: ({\it upper}) the Lomb-Scargle power spectrum, with dominant peaks at 1 cycle\,d$^{-1}$ and its harmonics, due to the data sampling window; ({\it middle}) power spectrum using CLEAN, with the strongest peak (arrowed) indicating the orbital ($P_{\rm orb}^{-1}$ = 1.4588\,d$^{-1}$)/superhump ($P_{\rm sh}^{-1}$ = 1.42241\,d$^{-1}$) frequencies; ({\it lower}) zoom into the CLEANed power spectrum strongest peak, the different coloured curves correspond to different time intervals as the outburst evolves (see inset box).  Note that for the first 86 days of the outburst, there is no significant power seen at all (the cyan curve very close to zero). From day 87 onwards, substantial power is detected at $P_{\rm sh}$ (green) and then close to $P_{\rm orb}$ (black), with the green and red dashed vertical lines indicating the relevant frequencies.
}
\end{figure}

While our initial Lomb-Scargle (LS) periodogram of the whole AAVSO 2018 light-curve (Figure \ref{fig:LS_AAVSO}, top panel) did reveal the $\sim$0.7d modulation described above, the power spectrum was dominated by large peaks at 1 cycle\,d$^{-1}$ and its harmonics. These were a result of the daily gaps introduced by the western hemisphere distribution of AAVSO observers, and produced strong features in the window spectrum. Accordingly we used the same approach as described in \cite{Barthes95} in order to effectively ``clean'' these spurious features from the power spectrum.  This is equivalent to pre-whitening the data \citep{Roberts87}, as used in a very similar analysis recently by \cite{Boyd17}.  We employed the CLEAN-PS code of \cite{Lehto93} \footnote{The clean algorithm performs a nonlinear deconvolution in the frequency domain and it provides a simple way to understand and remove the artifacts introduced by missing data.}, which is based on the algorithms of \citet{Roberts87} and implementation of \citet{Clark80}, in producing the overall power spectrum shown in the middle panel of Figure \ref{fig:LS_AAVSO}, now dominated by the $\sim$0.7d feature. {\color{magenta} } This is shown at much higher resolution in the lower panel, now revealing both the superhump and orbital periods.  We also divided these power spectra into 4 main time intervals of the outburst to show how these peaks evolved. 

\begin{table}
	\centering
	\caption{{\bf Optical Photometric Periods in J1820 (from Fig. \ref{fig:LS_AAVSO})}}
	\label{tab:Phot-P-tab}
\begin{tabular}{ c c c l}
\hline
\textbf{Day Nos.$^\dagger$} & \textbf{Freq.} (d$^{-1}$) & \textbf{P} (d)$^*$ & Notes \\
\hline
87 - 102 & 1.42241 & 0.70303(1) & $P_{\rm sh}$ \\
103 - 112 & 1.44797 & 0.69062(5) & \\
113 - 160 & 1.45301 & 0.68823(2) & $P_{\rm W}$\\
113 - 253 & 1.45301 & 0.68823(5) & $P_{\rm W}$\\
\hline
\end{tabular}
\begin{flushleft}
$^*$ 1$\sigma$ uncertainties in last digit given in parentheses.\\
$^\dagger$ These are the same time intervals as used in Figure \ref{fig:LS_AAVSO}.\\
Note that $P_{\rm orb}$ = 0.68549(1)\,d (freq. 1.4588\,d$^{-1}$) \citep{Torres19}. 
\end{flushleft}
\end{table}

We were not surprised to detect the $P_{\rm sh}$ signal at a frequency of 1.422\,d$^{-1}$ (=0.703 d, the green dashed line), as our AAVSO dataset incorporates the CBA (Center for Backyard Astrophysics, \citealt{Patterson13}) data of \cite{Patterson2018ATel11756} where it was first reported.  That $P_{\rm orb}$ is now well-defined spectroscopically means that this confirms the superhump signal as being a {\it positive} superhump, i.e. at a slightly longer period than $P_{\rm orb}$.

However, the $P_{\rm orb}$ signal has not been reported photometrically before. As indicated in Figure \ref{fig:LS_AAVSO}, a peak is found close to a frequency of 1.4588\,d$^{-1}$ (= 0.6855\,d, the red dashed line).  This corresponds to the spectroscopically-determined value of $P_{\rm orb}$ in \cite{Torres19}, although their precision in determining $P_{\rm orb}$ of $\pm$10$^{-5}$\,d, is much better than our $\sim$0.001\,d, due to their longer baseline.  This work also provides us with a precise ephemeris for our subsequent examination of the orbital phase-folded light-curves.

\cite{Patterson19} had already noted that they could see no periodic signals in their CBA data during the first $\sim$75 days of J1820's outburst, as demonstrated in Figure \ref{fig:LS_AAVSO} by the lowest curve, which covers days 10-86.  It is clear that the power spectrum changes dramatically as a function of time through the outburst.  

\subsubsection{Dynamical Power Spectrum analysis}

\begin{figure}
\centering
\includegraphics[width=20pc,height=19pc]{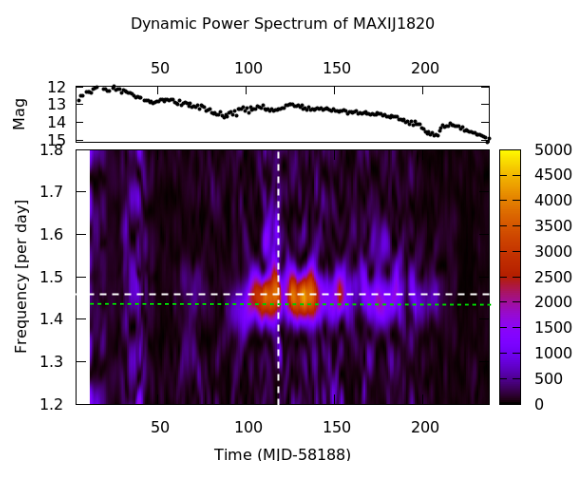}
\caption{\label{fig:DPS_Porb_Psh}
The AAVSO light curve (top panel) and dynamical power spectrum (DPS, see text), centred near the superhump and orbital frequencies, where the latter is marked by the dotted horizontal white line at 1.45879\,d$^{-1}$. Orbital power peaks at time $\sim$day 112 (MJD-58188), but just prior to this, the power is seen to be dominated at the lower (superhump) frequency (1.42241\,d$^{-1}$, the dotted green line). The vertical white dashed line marked at day 118 corresponds to the X-ray state transition time, which is reflected in the MAXI and NICER HR plots in Figure \ref{fig:LC}.}
\end{figure}

Accordingly we performed a Dynamical Power Spectrum (DPS) analysis on the well-sampled AAVSO dataset, extending from days 12 - 239, with the results shown in Figure \ref{fig:DPS_Porb_Psh}, which is centered around the frequency of $\sim$ 1.5\,d$^{-1}$. The DPS plot was produced using `sliding windows' that are 10d datasets (hence each window covers almost 14 binary cycles)
which move by 1d in the time domain to produce adequate resolution, and also provide a degree of smoothing of the resulting power spectra. Power is plotted here at the centre time for each 10d window. The horizontal white dotted line marks the exact value of the spectroscopically determined orbital frequency (1.45881\,d$^{-1}$), from the motion of the donor as measured in quiescence \citep{Torres19}. 

The superhump frequency of 1.422\,d$^{-1}$ (0.703 d) can be discerned in the DPS (Fig. \ref{fig:DPS_Porb_Psh}) as the enhancement of power, just below the dashed orbital frequency line, beginning at day 87, $\sim$15\,d before the orbital power peaks.  It is this interval that was first reported by \cite{Patterson2018ATel11756} and shown in more detail in \cite{Patterson19}, whose Figure 2 covers days 92-122 (which have the largest amplitude variability, and from which they derived a 0.69045 d period).

\subsubsection{The orbital and superhump light curves}

The power spectra (Figure \ref{fig:LS_AAVSO}) show that there is no periodicity present until day 87, when the superhump modulation appears and dominates the light-curve during the 15d interval from day 87-102 (Figure \ref{fig:LS_AAVSO}, bottom panel, green curve).   After day 102, the modulation has moved towards the orbital period (as can be seen in Figure \ref{fig:DPS_Porb_Psh}), where it remains strong, even after the state transition (day 118).  In fact, it is detectable throughout the subsequent soft state, up to the transition back to the hard state, around day 210, albeit with gradually decreasing amplitude. This is clear from Figure \ref{fig:LS_AAVSO}, bottom panel, in the red and violet curves, and where these key day numbers were used in selecting the intervals of interest. 
 
\begin{figure}
\includegraphics[width=0.5\textwidth]{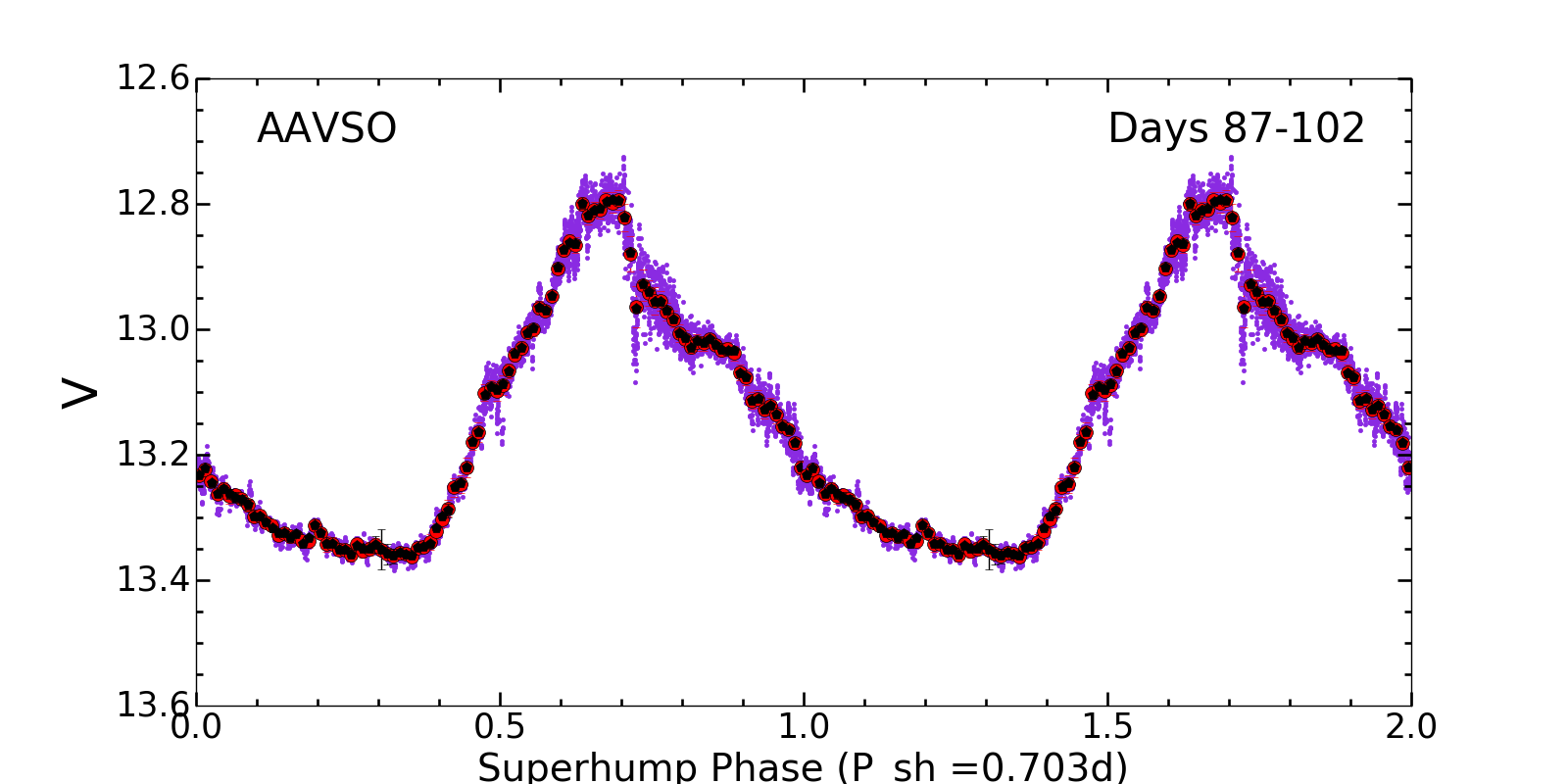}
\includegraphics[width=0.5\textwidth]{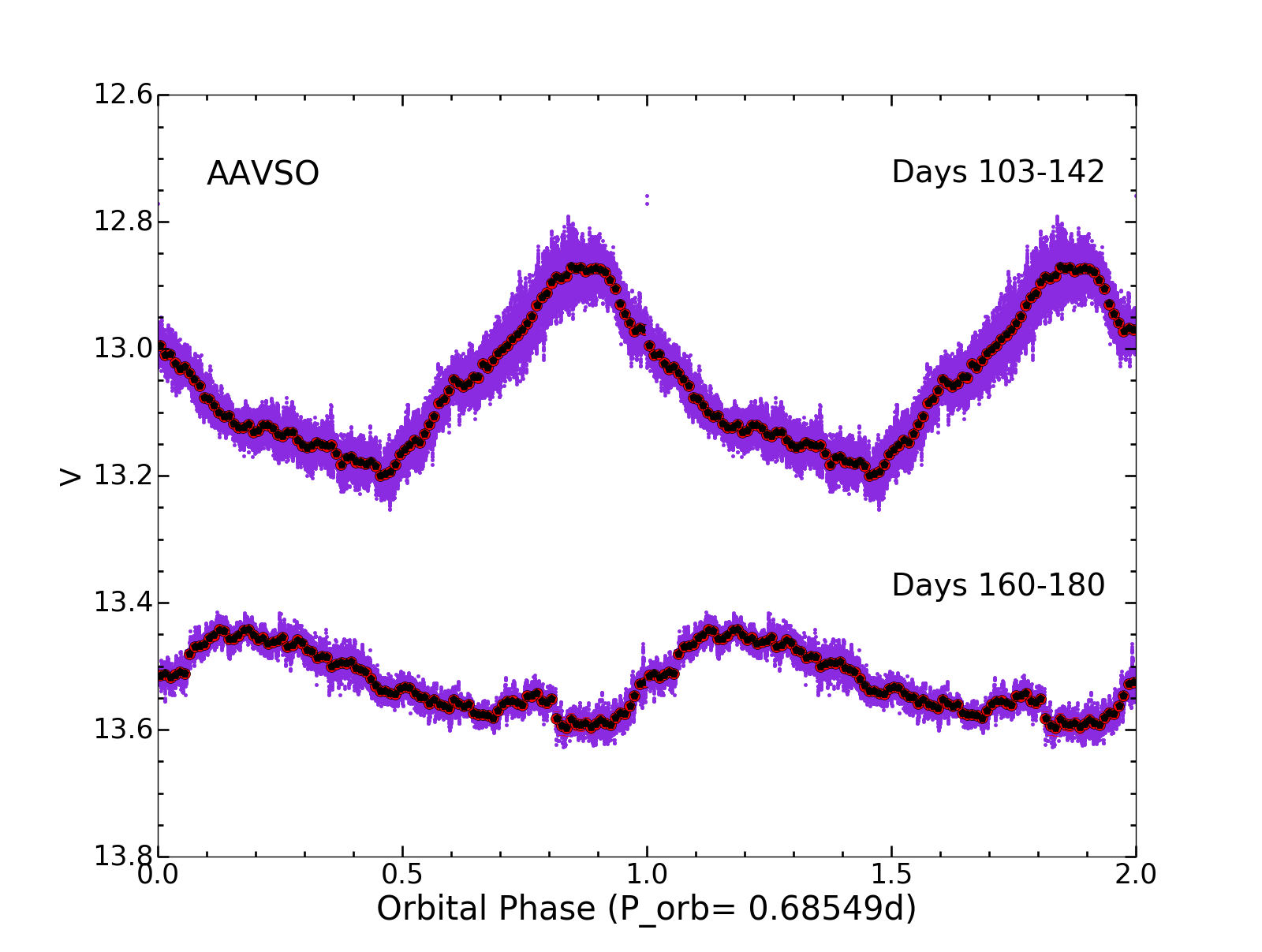}
\includegraphics[width=0.5\textwidth]{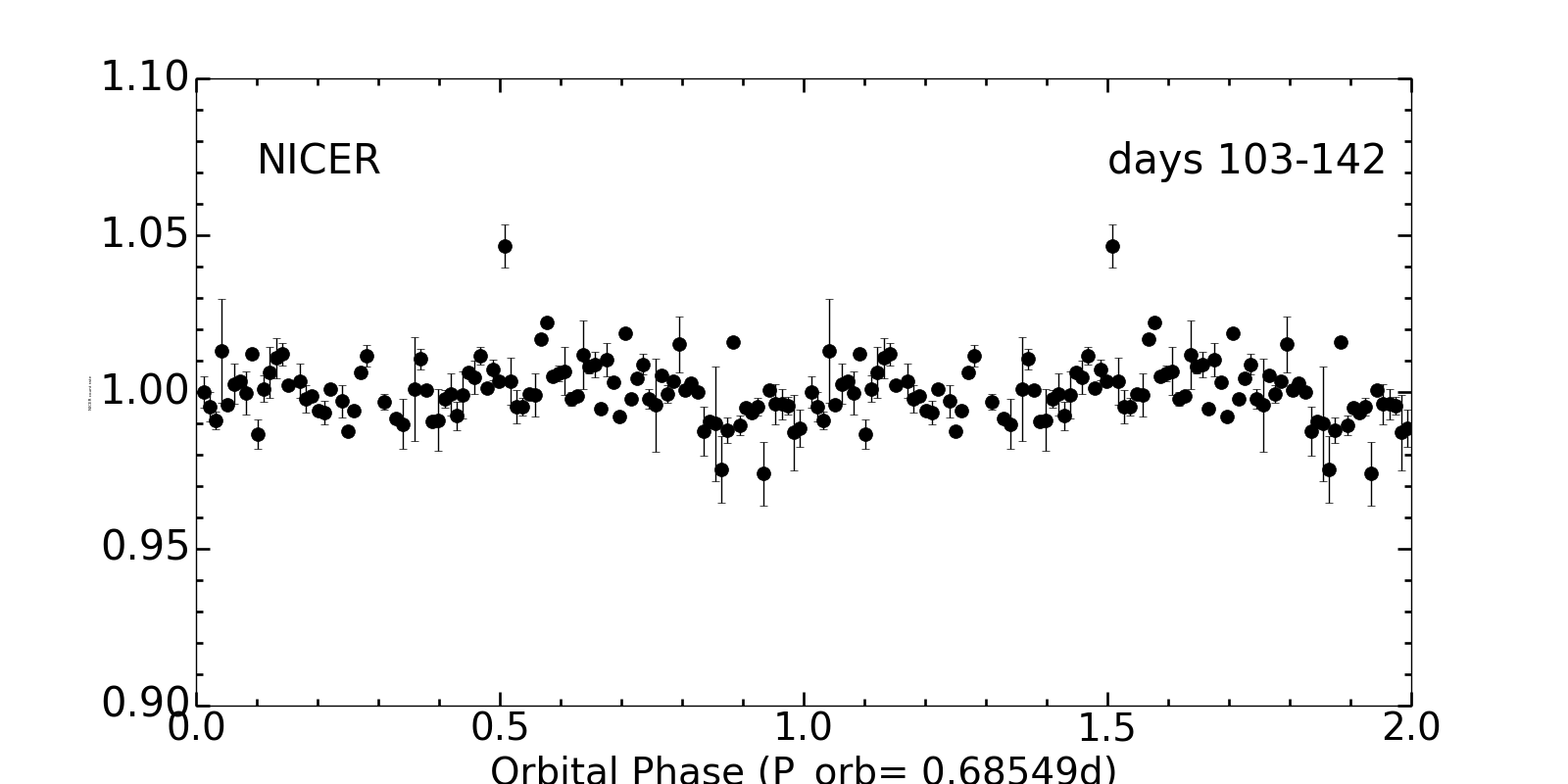}
\caption{\label{fig:phase_fold} 
Phase-folded AAVSO light curves on (top panel) superhump ($P_{\rm sh}$ = 0.703 days) and (middle panel) orbital ($P_{\rm orb}$ = 0.68549 days) periods, during the intervals where they dominate. The 50-bin means (and errors) are shown as solid black points.  The bottom panel is the normalised NICER 2-12\,keV flux also phase-folded on $P_{\rm orb}$ for days 103-142 where the optical modulation is at a maximum, showing that there is no X-ray equivalent. Note that $T_0$ for the orbital phase is taken from \citet{Torres19}, i.e. it is the spectroscopic ephemeris with phase 0 at inferior conjunction.}
\end{figure}

To investigate the evolution and variation in these periodicities at different epochs, we have folded the AAVSO data on both $P_{\rm orb}$ and $P_{\rm sh}$, as shown in Figure \ref{fig:phase_fold}. The top panel is the superhump light curve for days 87-102, showing the enormous ($\geq$0.6\,mag) amplitude present through this interval.  The second panel is a phase-fold on $P_{\rm orb}$ for days 103-142, the interval from Figure \ref{fig:DPS_Porb_Psh} where the strongest power close to $P_{\rm orb}$ occurs, and days 160-180, when it has diminished but is still present.  However, the key feature to note here is not just reduced modulation amplitude, but the movement in phase of the light-curve peak, from phase $\sim$0.9 to $\sim$0.2.  This will be investigated further in the next section.  We note also that \cite{Patterson19} reported a weak 0.6883 d modulation in their final 2 months of CBA data.

\begin{figure}
\includegraphics[width=0.5\textwidth]{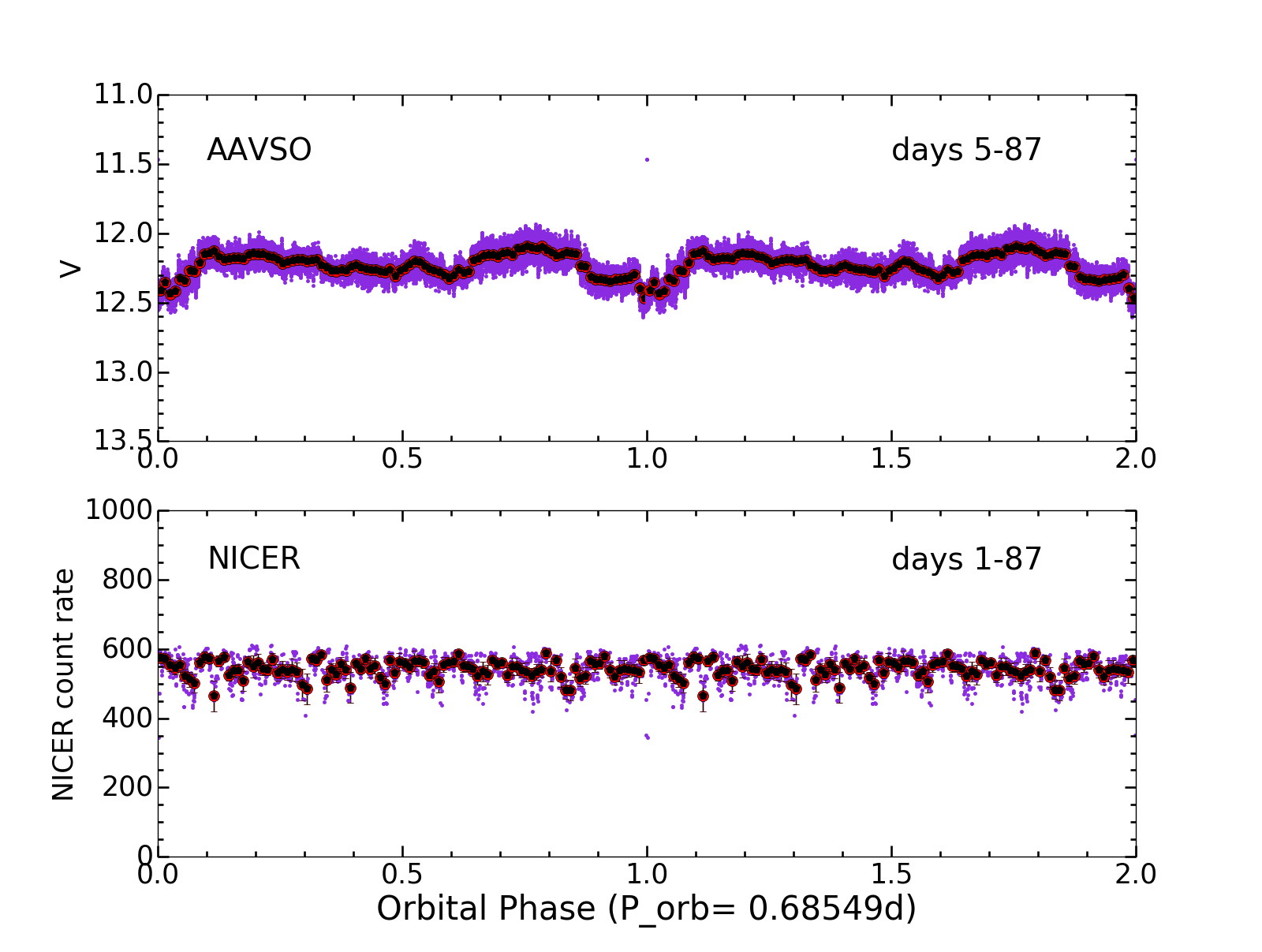}
\caption{\label{fig:0-87phase_fold} 
The AAVSO and NICER data for the first 87 days of outburst folded on $P_{\rm orb}$ using 100 phase bins per cycle.  There is no feature present in X-rays, but a possible weak partial eclipse is seen in the optical, although no signal is present in either Lomb-Scargle or phase-dispersion minimisation analysis.}
\end{figure}

It has been noted in most observational papers on J1820 that there was no orbital period signal present in the first 87 days of outburst.  However, given that fast ($\sim$minutes) X-ray dips were seen by XMM \citep{Kajava19}, this suggested that the light-curve might contain partial eclipses, dipping or other obscuration that would be highly non-sinusoidal and hence have been overlooked in previous studies.  Accordingly, we folded the first 87 days of AAVSO and NICER data on $P_{\rm orb}$, obtaining the light-curves shown in Figure \ref{fig:0-87phase_fold}.  There are no phase-dependent features evident in the X-ray light-curve, but the optical shows a possible weak ($\leq$0.2 mag) partial eclipse at phase 0, although it must be noted that there is no signal in either the Lomb-Scargle or Phase-Dispersion Minimisation periodogram analysis.

\begin{figure}
\includegraphics[width=22pc]{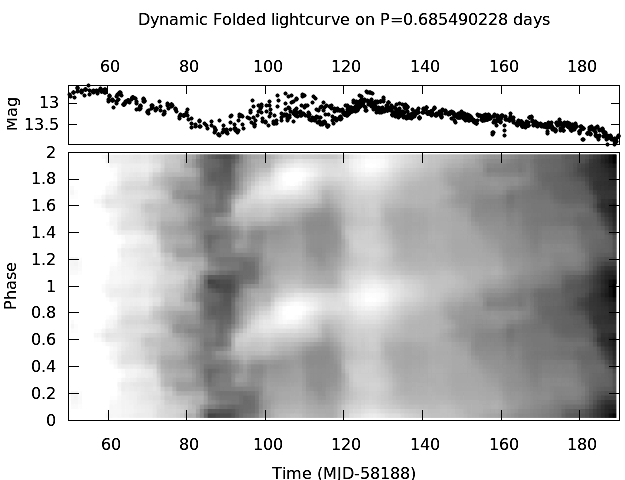}
\caption{\label{fig:Dynamic_fold} Dynamic fold of the AAVSO V-band data on the spectroscopically determined $P_{\rm orb}$ \citep{Torres19} during the time interval of days 50-190.  This makes it clear how little (orbital) modulation is present early in the outburst, but how dramatically the modulation begins at day 87, and then continues for the rest of the dataset.  The grey scale magnitude range is from 12.7 (white) to 13.8 (black).  See text for full details.
}
\end{figure}

\subsubsection{Dynamical Optical Light Curve}

With such high quality data, in which at certain times individual cycles of the superhump/orbital modulation were clearly visible, and given the phase shifts revealed in the previous section, we decided to perform a dynamic fold of the light curve itself.  This is shown in Figure \ref{fig:Dynamic_fold}, and  uses the same 10d-sliding window technique, with 1d steps, as described earlier, in which the data within that window are phase-binned on $P_{\rm orb}$ and the resulting light curve plotted as a grey-scale.  This covers the interval of days 82-162 in order  
to show how the modulation shifts with time, since Figure \ref{fig:DPS_Porb_Psh} has already indicated that a longer (than $P_{\rm orb}$) period is required during the first part of this interval, and a close to orbital modulation itself is dominant from day 102-142. 

Accordingly, a significant drift in phase of $\sim$0.3 is evident for the first $\sim$20d of Figure \ref{fig:Dynamic_fold}, which was also noted by \cite{Patterson19}. This is exactly the movement in (orbital) phase that would be expected for the $P_{\rm sh}$ = 0.703\,d that is present at this time in our power spectra (Fig. \ref{fig:LS_AAVSO}).  It should then be noted in Figure \ref{fig:Dynamic_fold} that, after $\sim$ day 103, the drift in the light-curve peak slows down, moving only by $\sim$0.25 in phase over the next 50 days.  This movement is shown even more clearly in the phase-folded light curves (Fig. \ref{fig:phase_fold}), which compares the orbital modulation for days 103-142 with 160-180.  This is the reason that the power spectra peaks for post-day 103 in Fig. \ref{fig:LS_AAVSO} are not {\it exactly} at $P_{\rm orb}$, but instead at a very slightly longer period that we denote $P_{\rm W} = 0.68823\,d$ (frequency of 1.45301\,$d^{-1}$), which is also marked on that figure.  These photometric periods detected are summarised in Table \ref{tab:Phot-P-tab}.

\subsubsection{The precession period}

\begin{figure}
\includegraphics[width=21pc]{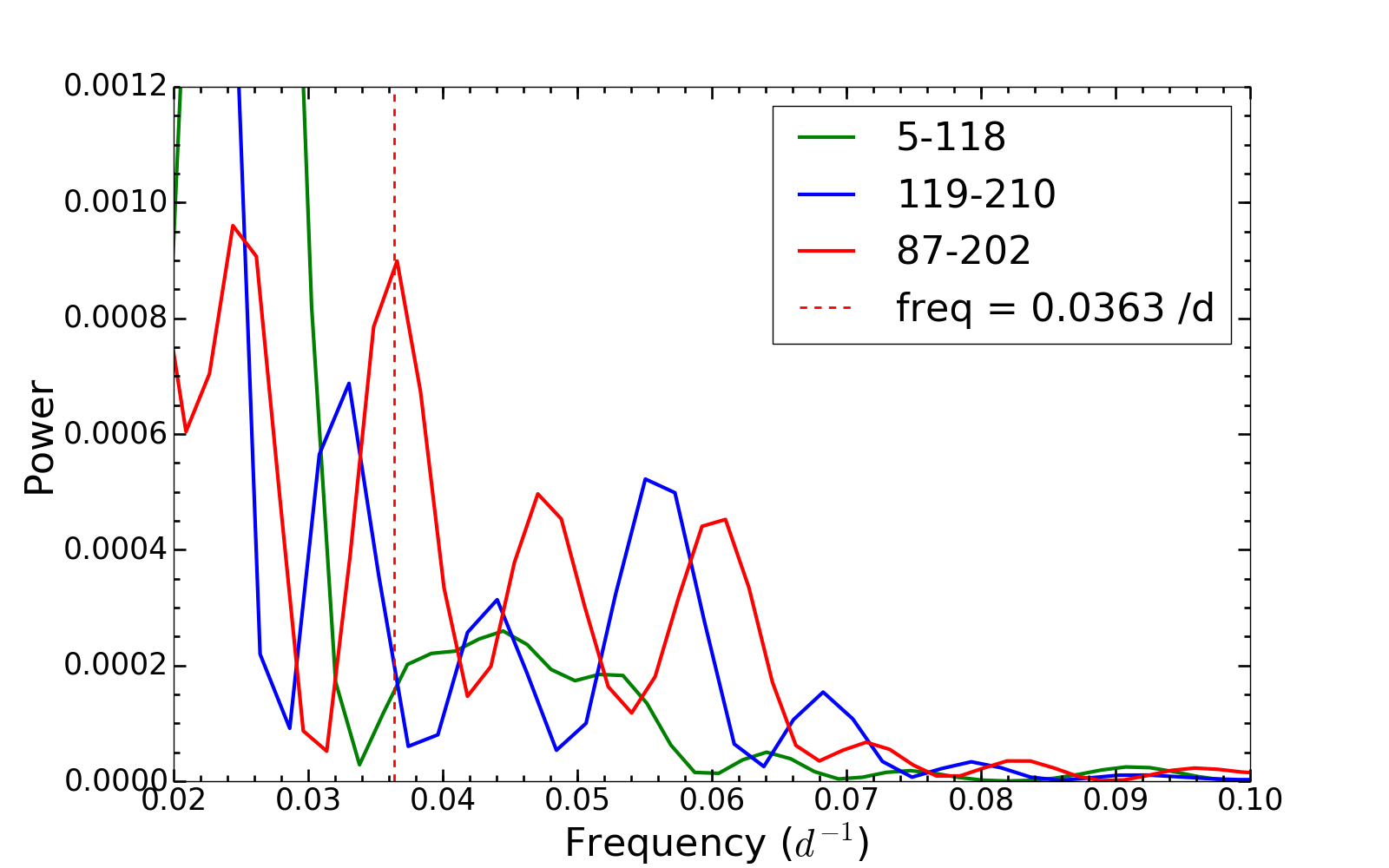}
\caption{\label{fig:DPS_precession} 
Zoom-in to the low-frequency end of Figure \ref{fig:LS_AAVSO}, covering the beat-period variations that might be expected between $P_{\rm orb}$ and $P_{\rm sh}$.} 
\end{figure}

With such an extensive dataset we had hoped to search for evidence of $P_{\rm prec}$ itself, something which is rarely possible.  The detection of a positive superhump in J1820 indicates that, in the commonly accepted model for superhumps (see Introduction, the disc is then precessing on the beat between $P_{\rm orb}$ and $P_{\rm sh}$, i.e. 27.548 d (or frequency of 0.0363\,d$^{-1}$).  This is comparable to the predicted value of $\sim$28 days in \cite{Torres19}. Such a low-frequency feature requires zooming-in to this part of Figure \ref{fig:LS_AAVSO} (middle panel) and this is shown in Figure \ref{fig:DPS_precession}.  While there is clearly substantial low-frequency power present, there is no stable signal present close to 0.0363\,d$^{-1}$ in either half of our dataset.  Nevertheless, the variability at these low frequencies ($\sim$20-30\,d) is directly visible in the overview AAVSO light-curve shown in the top panel of Figure \ref{fig:Dynamic_fold}.  It should also be noted that the drift in the orbital-folded light-curve discussed in the previous section (Figure \ref{fig:Dynamic_fold}), which results in the photometric $P_{\rm W}$, can be interpreted as a changed $P_{\rm sh}$.  This would then have a much longer ($\sim$175\,d) beat period with $P_{\rm orb}$.

\subsection{Optical/X-ray evolution through the outburst}

We now attempt to locate these large periodic variations exhibited by J1820 within the context of the overall optical/X-ray light-curve of Figure \ref{fig:LC}.  A key feature is the X-ray state change (from hard to soft) that occurs between days 110-120 (already noted by \cite{Stiele2020} for its unusual rapidity) and is clearly visible in the {\it Swift}-XRT, MAXI and NICER light-curves, as well as their hardness ratios. Not surprisingly, the quality and extent of the X-ray coverage through the transition has meant that the X-ray spectral and timing properties of these data have already undergone substantial analysis and interpretation, see e.g. \citet{Kara19}, \citet{Buisson19}, \citet{Paice19}, \citet{Stiele2020}, \citet{Wang20}, \citet{Paice_J1820Evolution_2021}, \citet{Axelsson21}, \citet{deMarco21}.  Our aim here is to build on these X-ray studies placing them in the context of our optical timing analysis.

\begin{figure}
\centering
\includegraphics[width=22pc,height=14.5pc]{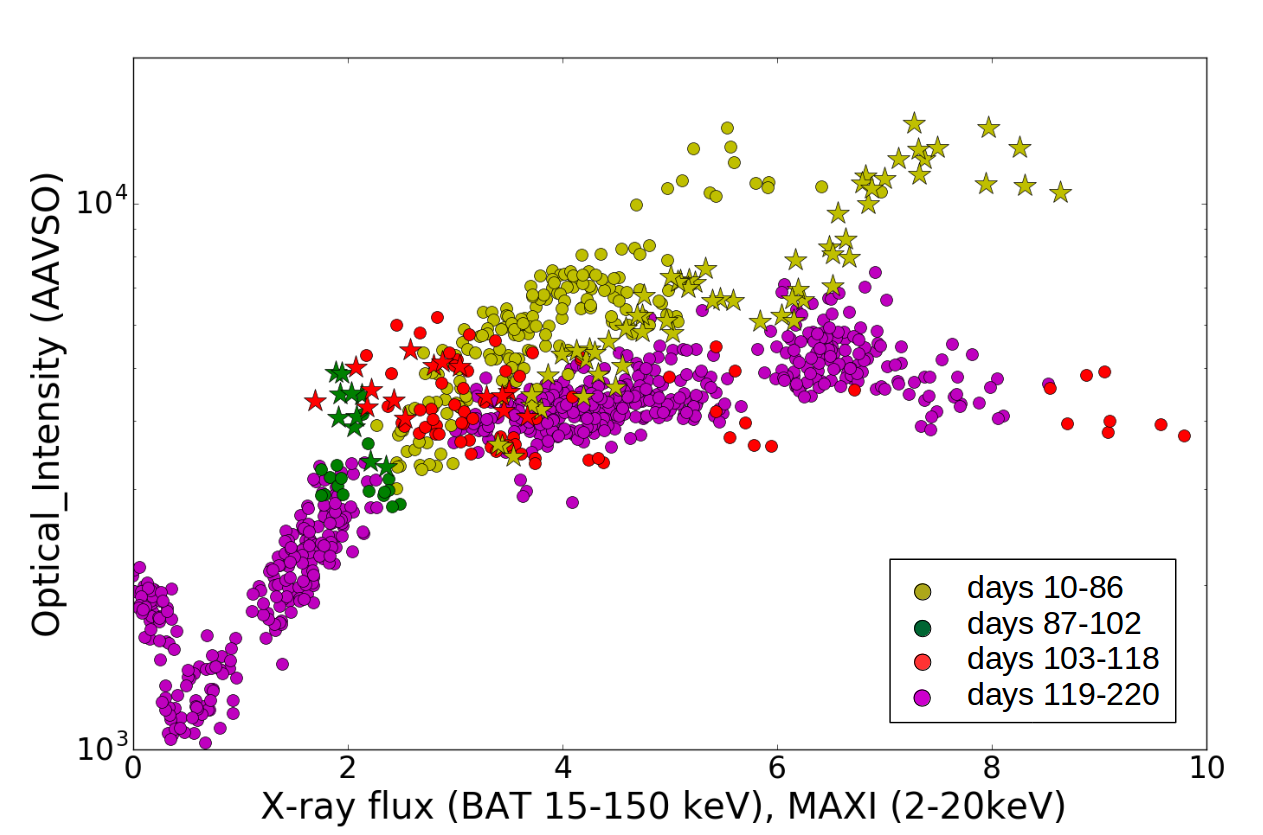}
\caption{\label{fig:optical_xray_flux} The optical versus MAXI (circles) and {\it Swift}/BAT (stars) X-ray fluxes for the different time intervals of the J1820 outburst, using the same colour scheme as in Figure \ref{fig:LS_AAVSO} lower box. i.e. we use four different colours for four different time intervals -- pale green covers the hard state, magenta points for the soft state, and the dark green and red points are when the largest optical periodic modulations were occurring (see text).}
\end{figure}

The state transition provides a natural division of the light-curve into intervals when its optical/X-ray behaviour has very different properties.  We demonstrate this by plotting the AAVSO V-band flux against the MAXI 2-20\,keV and {\it Swift}/BAT 15-150\,keV X-ray fluxes in Figure \ref{fig:optical_xray_flux}, where we use different colours to correspond to these key intervals.

From Figure \ref{fig:LC} it is clear that the AAVSO and MAXI light-curves follow the same basic shapes, on both sides of the state change, and this overall correlation is clear in Figure \ref{fig:optical_xray_flux}.  We have included the {\it Swift}/BAT data in this figure (star symbols) to demonstrate that the optical follows the hard X-ray flux even more closely.  This is largely as expected in an LMXB, where the optical flux is driven by an X-ray irradiated disc \citep{JVP94}.  However, the two states have different slopes in this correlation, with the soft state (magenta) being much flatter than the hard state data.  Note that the hard state, which lasts until the transition begins around day 112, is divided into 3 intervals, using the same colour scheme as in our power spectral analysis (Figure \ref{fig:LS_AAVSO} lower box). This is in order to carefully examine any X-ray variability that might be associated with the intervals (days 87-102 and 103-142) when the largest optical modulations are present.

Even though the optical flux from luminous LMXBs has been clearly demonstrated to arise from X-ray irradiation of the disc and inner face of the donor, it is clear that the large modulations present during days 87-102 (dark green symbols) are {\it not} driven by X-ray variations.  We demonstrate the absence of any X-ray modulation on $P_{\rm orb}$ in Figure \ref{fig:phase_fold} (bottom panel) where the higher time-resolution NICER data is phase-folded.  

It is also interesting to note that the slight increase in the optical at very low X-ray fluxes, relative to the general trend, is actually the ``bump'' in the AAVSO light-curve around day 210, and has no equivalent in the MAXI light-curve.

\subsubsection{X-ray State Transition} 
An important feature of Figure \ref{fig:LC} is that the optical flux stops its gradual decline from outburst and increases again, starting at day 87, which is exactly when the $P_{\rm sh}$ power starts to appear (Figure \ref{fig:DPS_Porb_Psh}), and just before the {\it Swift}/BAT begins its rise to its secondary maximum.  Yet J1820 remains in the hard state for almost another month, the interval during which the periodic optical modulations are strongest. The amplitude of these modulations drops significantly at the end of this interval, which is when the X-ray state transition occurs and there is a sudden rise in X-ray brightness detected by MAXI, Swift and NICER.  This close correlation indicates that the optical and X-ray behaviour {\it must} be physically linked.

After the transition, J1820 remains in the soft state for $\sim$80 days, until $\sim$day 212, when it changes back to the hard state.  These are the magenta points in Figure \ref{fig:optical_xray_flux}.

\begin{figure}
\centering
\includegraphics[width=0.5\textwidth]{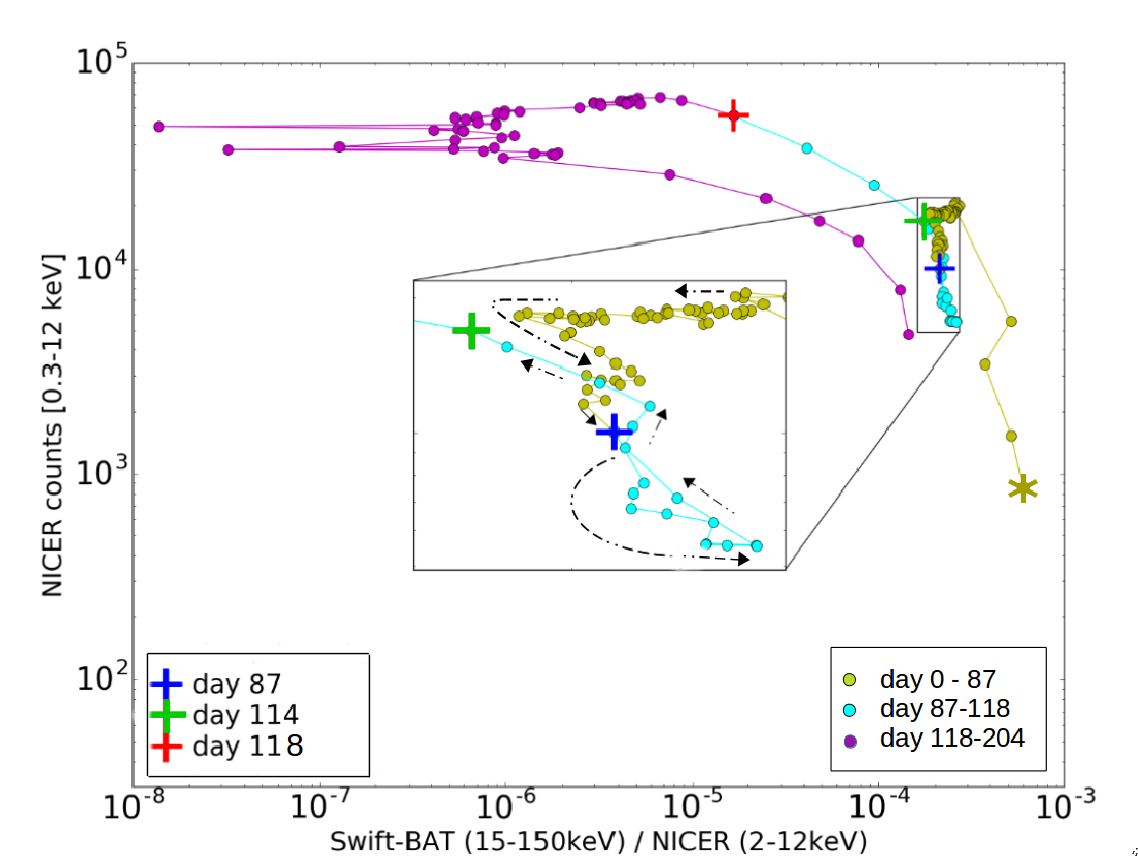}
\caption{\label{fig:Luminosity}HID for J1820 using BAT(15-150keV)/NICER(2-12keV) to define the hard colour, and covering the first 200 days after outburst (marked as ``*''), and colour-coded as: days 0-87 (pale green), 87-118 (cyan), 118-204 (magenta).  The crosses mark (blue, day 87) the start of large optical modulations, (green, day 114) the BAT peak, and (red, day 118) the radio flare/jet ejection.  The state transition occurs between days 115-120.  The interval of large optical modulation is plotted in cyan, and blown up (inset) to show more detail, with the direction of time marked by the arrows.  }
\end{figure}

The X-ray spectral changes that take place during the outbursts of BHXBs are usually examined via the Hardness-Intensity Diagram (HID), and these have been presented in many of the papers referred to earlier in this section.  However, given the close relationship of the {\it Swift}/BAT and optical light-curves, we have created our own HID (Figure \ref{fig:Luminosity}), using the NICER total counts for the intensity and the {\it Swift}/BAT 15-150\,keV/NICER 2-12\,keV ratio as the ``hard colour''.  This plot covers the entire hard and soft states (220 days) of the outburst, and we use essentially the same colour scheme as in earlier figures, so as to facilitate cross-referencing.  

There are 3 key times leading up to the state transition around day 120, and these are marked on our HID as crosses, and on Figure \ref{fig:LC} as dashed lines:
\begin{itemize}
    \item Blue cross - day 87: the start of the large optical modulation
    \item Green cross - day 114: the peak of the {\it Swift}/BAT secondary maximum
    \item Red cross - day 118: the radio flare/jet ejection \citep{Bright2020,Homan20}
\end{itemize}
Our HID shows that there are rapid and complex changes occurring during this $\sim$30d interval, so the inset in Figure \ref{fig:Luminosity} shows a zoom-in, where arrows indicate the direction of time.  It is particularly interesting that rapid X-ray spectral changes begin on day 87 (just as the optical modulation appears, at the blue cross), as the source enters a loop (marked in cyan for clarity), becoming first harder and then softer as it brightens towards the BAT peak (at the green cross).  It then continues to soften as it moves rapidly (within a few days) to the time of the AMI radio flare and superluminal jet ejection episode at the red cross \citep{Bright2020}, which essentially marks the state transition.  The optical modulation continues into the soft state (as can be seen in the DPS, Figure \ref{fig:DPS_Porb_Psh}), remaining strong for another month, and while its amplitude gradually fades after that, it is still at almost 0.2\,mag (peak-to-peak) in the days 160-180 light-curve of Figure \ref{fig:phase_fold}.  This remarkable correlation between the optical and X-ray behaviour must have a physical connection that we will explore in the next section.

\section{Discussion}\label{sec:5}

Our analysis of the extensive AAVSO optical and {\it Swift}/NICER light-curves of J1820's 2018 outburst shows 3 main phases, with the following properties:

\begin{enumerate}

\item {Days 0-86:  classic BHXB X-ray/optical outburst light curve in the hard state, decaying and variable, but with no  (substantial) periodic modulation.  Note also the change in NICER flux slope at $\sim$day 60 in Figure \ref{fig:LC}.}
\item {Days 87-112: optical decline is reversed and a gradual increase in brightness occurs, along with the sudden appearance of a huge optical modulation and a secondary maximum in {\it Swift}/BAT.  The modulation is first on $P_{\rm sh}$, then drifts to be very close to $P_{\rm orb}$, but measurably different at $P_W$=0.68823\,d, and with an amplitude at times $\geq$0.5mag. }
\item{Days 112-253: begins with an X-ray state change from hard to soft.  Optical modulation continues at  $P_W$, but with a slowly decaying amplitude.}

\end{enumerate}

\begin{figure}
\centering
\includegraphics[width=0.5\textwidth]{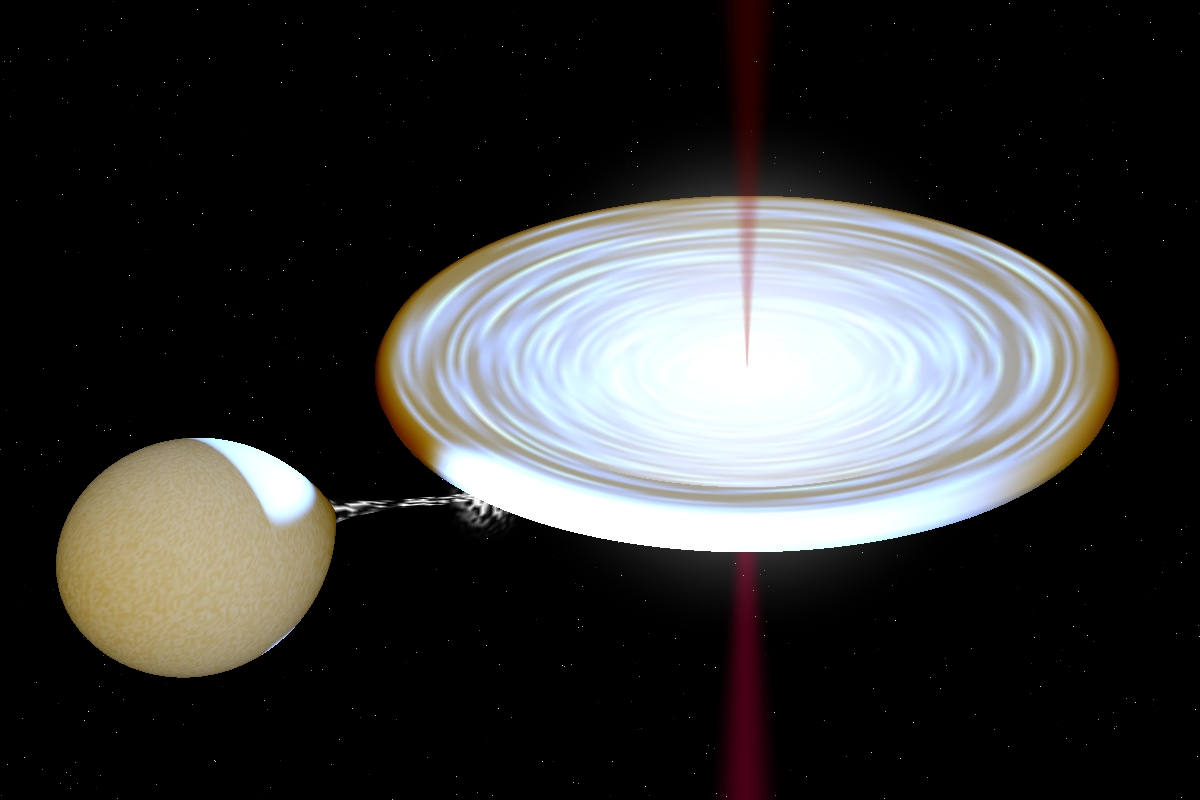}
\caption{\label{fig:Phase0.85visual} Visualisation of the J1820 system at orbital phase 0.85, using BinSim$^{4}$ and the binary parameters given in the text.}

\footnotesize{\textit{$^4$}} \url{http://www.phys.lsu.edu/~rih/binsim/download.html}\\
\end{figure}

Of these remarkable properties, the principal one is the large amplitude (mean $\sim$0.6 mag) modulation on $P_{\rm sh}$ (Figure \ref{fig:phase_fold} top panel), which has never been seen before in the BH XRTs, and appears at the same time as the {\it Swift}/BAT light curve starts to increase.  Indeed, \cite{Patterson19} already named J1820 as the ``king of the black-hole superhumps'', which is an apt title when it is compared (Table \ref{tab:XRT-SHs-table}) with other LMXB BH X-ray transients observed during outburst, as their superhump modulation amplitude, $A$, is always $\leq$0.2 mag.  Furthermore, our analysis of this modulation during the soft state, when it is occurring at $P_{\rm W}$, has shown that the peak in the folded light-curve (see Figure \ref{fig:Dynamic_fold}) moves gradually during the next 80\,d from orbital phase $\sim$0.85, which is close to inferior conjunction of the mass-losing companion (with an orientation similar to that visualised in Figure \ref{fig:Phase0.85visual}) to phase $\sim$0.2.  This means that the optical modulation {\it cannot} arise on the X-ray irradiated face of the donor, and must instead somehow be a feature of the disc itself.

To understand this property, we first need to examine the background to superhumps in cataclysmic variables (CVs), how this has been applied to LMXBs, and how this is, or is not, relevant to J1820.

\begin{table*}
	\centering
	\caption{Black-hole X-ray transients displaying superhumps during outburst}
	\label{tab:XRT-SHs-table}
\begin{tabular}{l c c c c c}
\hline
\textbf{System name} & \textbf{P$_{\mathrm{sh}}$} (d)& \textbf{P$_{\mathrm{orb}}$} (d) & \textbf{$\epsilon$} (\%)  & \textbf{A} (mag) & \textbf{i} (deg)\\
\hline
XTE~J1118+480$^1$ & 0.169930 & 0.170529 & 0.35 & 0.1 &71--82\\
GRO~J0422+32 $^1$ & 0.2157 & 0.21216 & 1.67 & 0.1 & 10--50\\
GS~2000+25$^1$ & 0.3474 & 0.344098 & 0.96 & 0.2 & 54--60\\
GRS~1124-683$^1$ & 0.4376  & 0.4333 & 0.99 & 0.2 & 39--65\\
MAXI~J1820+070$^2$ & 0.70303 & 0.68549 & 2.6 & 0.36 & 66--81\\
\hline
Swift~J1753.5-0127$^3$ & 0.1351 & ? & ? & 0.15 & ?\\
GRS~1716-249$^4$ & 0.6127 & ? & ? & 0.1 & ?\\
\hline
\end{tabular}
\begin{flushleft}
$^{1}$\cite{Uemura04}; $^{2}$ this work;
$^{3}$\cite{Zurita08};
$^{4}$\cite{Masetti96}; 
\end{flushleft}
\end{table*}

\subsection{Superhumps and precessing accretion discs}

The SU UMa, short $P_{\rm orb}$ (mostly $<$2h), sub-class of CVs are where the {\it superhump} phenomenon was first encountered, so-called because they only occur during the extended ``superoutbursts'' of these systems, which otherwise display normal outbursts as seen in many dwarf novae (see e.g. \citealt{Warner95} for a thorough review).  While the ``humps'' in the light curves of many dwarf novae during outburst occur on $P_{\rm orb}$ (since they are a manifestation of the energy released when the mass-transfer stream impacts the accretion disc), superhumps have the property that they occur on a period that is always a few \% longer than $P_{\rm orb}$, and the period excess $\epsilon$ ($= (P_{\rm sh}-P_{\rm orb})/P_{\rm orb}$) is related to $q$ through the relation 

\begin{equation}
    \epsilon = 0.18q + 0.29q^2
\end{equation}

derived from observations of many SU UMa systems \citep{patterson2005}.

The mass donors in SU UMa systems have very low masses ($\leq$0.3M$_\odot$), and hence correspondingly low mass ratios, with ${q = M_2/M_{\rm WD}} \lesssim$ 0.25, and it is this feature that is a key part of what has been the common superhump explanation. It is also why they are often considered to be an excellent analogue of the BH X-ray transients, amongst other LMXBs.  Superoutbursts in SU UMa systems always follow a normal outburst, and this was explained by \cite{WhitehurstKing1991} as the hot accretion disc expanding towards its tidal radius, becoming eccentric and beginning to slowly precess as it reaches the 3:1 resonance.  They show that this is only possible if \textit{q} $\lesssim$ 0.3.  During closest approaches to the donor, tidal stresses increase, leading to greater heating of the disc and it becomes brighter, which we see as superhumps, on a period that is the beat between the precession and orbital periods.

It is this extreme $q$ value that completes the analogy with BH X-ray transients (BH-XRTs).  \cite{Torres2020} have already accurately measured $q$ spectroscopically for J1820, and it is 0.072 ($\pm$0.012).  However, the crucial difference is that, in CVs, the dominant optical light source is the thermal, viscous energy release within the disc itself, whereas in LMXBs it is the intense X-ray irradiation from within the inner disc illuminating the outer disc that provides the principal energy source, typically exceeding the disc's intrinsic energy by a factor ${\sim}10^3$.  Consequently, superhumps were not expected to be seen in BH-XRTs, and yet \cite{O'DonoghueCharles1996}'s careful reanalysis of the outburst photometry of 3 XRTs clearly showed that superhumps {\it were} present (and more were  subsequently discovered, as listed in table \ref{tab:XRT-SHs-table}).  This paradox seemed to be resolved by \cite{Haswell2001} who showed that, in precessing disc simulations, the disc area also varied, and this variation would therefore be reflected through the X-ray irradiation in the resulting optical light-curve.

\subsection{Application to MAXI J1820+070}

\subsubsection{Basic parameters}
Our observations show that, for J1820, $\epsilon$ = 0.026, which  would imply a mass ratio of $q$=0.12 (from equation (1)).  Interestingly, this is exactly the same as that first estimated by \cite{Torres19}, but still consistent, within the uncertainties of $\epsilon$, with the more accurate value of \cite{Torres2020}.  

Of course, J1820 has a much longer $P_{\rm orb}$ than any SU UMa system, so we can use the already established binary parameters to examine the scale of the binary.  From \cite{Torres2020} and Kepler's 3rd Law, the binary separation is $a = 6.8 R_\odot$, and, using the \cite{Eggleton83} formula, then the BH Roche lobe radius is $R_X = 4.1 R_\odot$.  For the important disc radii we use the approximations of \cite{Gilfanov05} for the circularisation and tidal radii:

\begin{equation}
    \frac{R_{\rm circ}}{a} = 0.074 \left(\frac{1+q}{q^2}\right)^{0.24}
 \end{equation}
 \begin{equation}
    \frac{R_{\rm tid}}{a} = 0.112 + \frac{0.270}{1+q} + \frac{0.239}{(1+q)^2}
\end{equation}

where both are accurate to $\sim$3\% over the range {\bf 0.03 (for $R_{\rm circ}$) or 0.06 (for $R_{\rm tid}$)} ${\leq}q{\leq}10$.  For J1820 these give $R_{\rm circ}$ = 1.9 $R_\odot$ and $R_{\rm tid}$ = 4 $R_\odot$, while the 3:1 tidal resonance is at \citep{WhitehurstKing1991}

\begin{equation}
    \frac{R_{\rm jk}}{a} = \left(\frac{\rm j-k}{\rm j}\right)^{2/3} (1+q)^{-1/3}
\end{equation}

which, for $\rm j=3, k=2$, gives $R_{32}$ = 3.2 $R_\odot$, hence making J1820 potentially susceptible to this resonance.

\subsubsection{Location and extent of the optical modulation}

At first sight, the optical modulations observed during the outburst of J1820 seem to follow the description and predictions of \cite{O'DonoghueCharles1996}
and \cite{Haswell2001}. The former note that superhumps are most likely detectable two months or more after the outburst beginning and their appearance
seems to be associated with a ``glitch'' in the (hard) X-ray lightcurve, which is exactly what is happening with J1820. They also point out that, contrary to CV superoutbursts,
in X-ray transients one observes during outbursts both the superhump and orbital modulations in the system's optical emission, as appears at first sight to be the case for J1820.

\cite{Haswell2001} note that the superhump tidal--resonance model cannot apply as such to the BH XRTs, since the supposedly tidally--enhanced viscous dissipation
is negligible compared to the contribution by X-ray irradiation, the ratio of the two corresponding luminosities being as low as $\sim 10^{-4}$. They suggest therefore that
the observed superhump modulation results from the varying disc surface area generated by the tidal--resonance between the disc and the secondary, as described at the beginning of this section. This 
would change the area visible to the observer, thereby causing modulations in the optical flux. This type of optical modulation would dominate in the outbursts of low--inclination binaries,
while in the higher--inclination systems a modulation at the orbital period, resulting from X--ray irradiation of the secondary, would be more pronounced.

At least two properties of the J1820 outburst are, however, incompatible with the above--described picture. First, the superhump amplitude reaches 0.5\,mag or greater, while the \cite{Haswell2001}
model provides for 0.1\,mag at most, since in the simulations on which it is based, the disc surface area changes by no more than about 10\%.\footnote{Also in the case of CV superhumps, for probably different reasons, the tidal--resonance model fails to reproduce the observed superhump amplitudes \citep[see][]{Smak09}}. Second, the optical modulation, because of its phasing cannot be attributed to the irradiated face of the secondary.  Such detailed examination of the phasing of the optical light in the other SXTs in Table \ref{tab:XRT-SHs-table} has not been possible, and an alternative explanation may therefore be needed for all of them.

Another key fact provided by J1820 is that its high orbital inclination, combined with lack of any X-ray modulation on this period, does require that we seek an explanation associated with the properties of the disc.  Furthermore, we have the unique result here that the modulation only begins at a particular time during the X-ray spectral evolution of the source. Given the significance of J1820's inclination in this discussion, we will first revisit the current observational constraints on $i$.  These are collected together in Table \ref{tab:J1820-i-tab}.

\begin{table}{h}
	\centering
	\caption{{\bf Inclination Measurements for J1820}}
	\label{tab:J1820-i-tab}
\begin{tabular}{ l c l l}
\hline
\textbf{Method} & \textbf{$i$ (${\circ}$)} & \textbf{Reference} & Notes \\
\hline
Opt. spec. & $>$69; $<$77 & \citet{Torres19} & H$\alpha$ EW; no eclipse \\
Opt. spec. & 66 -- 81 & \citet{Torres2020} & $v_{\rm rot}$\\
Radio ejecta & 63$\pm$3 & \citet{PAtri2020} & Jet axis \\
X-ray light-curve & $\sim$60 & \citet{Kajava19} & X-ray dips \\
Opt. phot. & 60 -- 70 & this paper & Partial eclipse \\
\hline
\end{tabular}
\end{table}

The absence of X-ray eclipses provides a strong constraint of $i <$77$^\circ$. But the initial indication of a value close to 70$^\circ$ came from \citet{Torres19} who interpreted their H$\alpha$ EW light-curve, peaking near orbital phase 0.9, as an indication of a grazing eclipse.  Suggestions that it might be lower came from \citet{PAtri2020}, whose radio ejecta defined the jet axis of the rotating BH to be $i$ = 63$\pm$3$^\circ$, and assuming that the jet axis is perpendicular to the plane of the accretion disc. Similarly lower values are supported by the X-ray dips (\citet{Homan18}, \citet{Kajava19}) and the possible optical partial eclipse seen in Figure \ref{fig:0-87phase_fold}.  We also suggest that the H$\alpha$ behaviour seen by \citet{Torres19} is actually due to the stream-impact hot spot, as might be expected during the first quiescence interval of J1820, as a consequence of ongoing mass-transfer immediately before a subsequent new outburst.  Accordingly, we adopt a value of $i$ = 63$^\circ$ for the remainder of this discussion, as visualised in Figure \ref{fig:Phase0.85visual}.

\subsubsection{A precessing, warped disc}

\begin{figure}
\begin{center}
\includegraphics[width=\columnwidth]{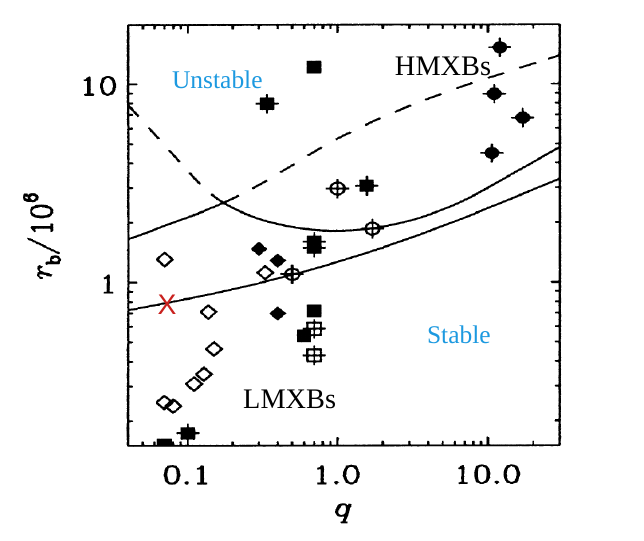}
\caption{{\bf \citet{Ogilvie01} computed the stability of X-ray binaries to irradiation-driven warping and this is their Figure 7, on which we have added J1820 as a red cross.  This plots the size of the binary, $r_b$, in units of $GM_1/c^2$, against $q$, and shows the unstable and stable regions, with HMXBs in the upper right, and LMXBs in the lower left.  J1820 is almost exactly on the stability curve for mode 1 warping (see text for details).  The source directly above it is V404 Cyg.}}
\label{fig:XRBstability}
\end{center}
\end{figure}

The solution to the superhump amplitude difficulty can also be found in \cite{Haswell2001} who invoke the possibility of the disc in X-ray transients being warped under radiation-induced torques 
\citep{Pringle96,Wijers99,Ogilvie01}. Indeed, as seen from Figure \ref{fig:XRBstability} , with $r_b/10^6 \sim 0.5$ ($r_b$ being the orbital separation) and $q=0.06$, J1820 is close to being 
unstable to the mode 1 (prograde) warping induced by radiative torques (and which assumes that $\alpha$=0.3 and $\epsilon$=0.1, see below). 

Since the large superhump amplitude cannot be produced by the disc's tidal deformation, but might be easily accounted for by its warped surface (in providing a much larger area for X-ray irradiation), the precession period should be identified with the warp precessional movement, i.e. with nodal precession. Unfortunately, the theory of radiatively-warped discs is not developed enough to produce reliable estimates of the nodal precession period resulting from this mechanism. One should also keep in mind that the mechanism's stability criterion, i.e. its critical radius, strongly depends on the viscosity parameter ($r_{\rm crit} \sim \alpha^{-4}$) and accretion radiative efficiency ($r_{\rm crit} \sim \epsilon^{-2}$). While the viscosity parameter in an irradiated disc is supposed to be $ > 0.1$ \citep{Tetarenko18} and is not expected to vary, the radiative efficiency clearly does vary with the X-ray spectral state of the system, so changes in the warp structure during outburst should be expected (see below).

Interestingly, the $P_{\rm sh}$ observed  in J1820 corresponds very well to the prediction of the 3:1 tidal model for a free--particle disc. The significance of this fact is unclear, especially given that superhumps in most SU UMa stars \textsl{do not} satisfy this relation \citep{Pearson06,Smith07,Smak20}.

The characteristic time for the growth of the 3:1 instability is \citep{Lubow91}
\begin{equation}
\tau \sim 0.08\, C^{-1} \, P_{\mathrm{orb}} q^{-2} 
\end{equation}
where C is a parameter taking into account the size of the disc, and $C=1$ for a ring (for which the calculations are done, see also \citet{goodchild06}).
For J1820 $\tau \sim 15 C^{-1}$ days, that requires a (probably) unrealistic $C$. It is therefore not clear if the 3:1 tidal-resonance plays any role in
the appearance of the superhump at day 87 of the J1820 outburst.

The strong, close to orbital modulation, $P_W$, appearing after day 103 must be related to the stream impact interactions with the outer disc regions, since the stream trajectory is the only
structure fixed in the orbital frame, apart from the secondary that must be excluded because of the slowly changing orbital phase of the modulation.  Furthermore, the ``hot spot'' is 
unlikely to be the source of the observed optical radiation. The reason is similar to that mentioned earlier in excluding the disc optical emission to be due to viscous dissipation, since
in both cases it is the local gravitational energy that is released. The hot spot luminosity is 
\begin{equation}
\label{eq:lspot}
L_{\rm spot} \approx 4 \times 10^{32}\left(\frac{\dot M}{10^{17}\,\rm g\,s^{-1}}\right) \rm erg\,s^{-1},
\end{equation}
so a mass-transfer rate $ \gtrsim 10^{18}$ g/s would be needed to account for the observed optical flux. However, such a mass-transfer rate is larger
than the critical accretion rate above which the disc is unstable \citep{Hameury20}, so that a rapid enhanced mass-transfer rate would at least modify the X-ray lightcurve, thereby
stopping the outburst's X-ray decay since the viscous propagation time of a mass-excess with width $\Delta R$ is
\begin{equation}
\label{eq:tvisc}
t_{\rm vis}\approx 70\, \alpha^{-1}_{0.2}R_{11}^{1/2}M_1^{1/2}T_4^{-2} \frac{\Delta R}{R}\, \rm days,
\end{equation}
but no such effect has been observed. The X-ray spectral event at day 118 is clearly a state transition, most probably due to the inner disc radius reaching the innermost stable
circular orbit.

Even if the stream impact cannot provide the luminosity required to explain the observed optical modulated flux, the stream must play a role in its appearance.
As explained in \citet{Ogilvie01} the character of the radiative-warping instability depends strongly on the radius at which matter is added to the disc. Even 
in the case of a planar disc, a substantial fraction of the mass-transfer stream can over(under)flow its surface, adding mass at the circularisation radius $R_{\rm circ}$
\citep[e.g.,][]{Armitage98}, but in the case of a warp most of the stream can end up at this radius, depending on the phase of the precession.
When mass is added to the external disc edge, the prograde mode 1 is becoming unstable, as should be the case for J1820. In addition, the critical
radius at which mode 1 is unstable is smaller than when mass is added at the circularisation radius. The calculations of \citet{Ogilvie01} do not take into account the fact that the mass added at the outer-disc radius has lower angular momentum than the disc at this  radius. These authors note that this produces an extra torque that might affect the stability
properties. They speculated that a disc could be unstable to radiation-driven warping when mass is added at the outer disc radius, but stable when it
is added at the circularization radius, so that such a system would display warping cycles: an initially flat disc with mass input at its
outer edge becomes unstable and tilts; mass input then moves towards the circularisation radius where the disc then becomes stable to warping and
resumes its initially flat shape. They do not consider modulations of the mass--transfer rates that could produce a similar effect if the amount of matter overflowing the disc
depends on the rate at which it is provided.  We stress that all these ``scenarios'' are very uncertain because of the absence of accurate calculations 
of the radiation-driven warping taking into account all the relevant torques and mass input to the disc. Nevertheless, it seems that J1820 during outburst is a system
where at least some of the processes conjectured by \citet{Ogilvie01} are occurring.

We can then suggest the following scenario based on the assumption that according to the \citet{Ogilvie01} criterion, J1820 during its outburst becomes unstable
to radiative warping, and taking into account our observation that a $P_{sh}$ to $P_W$ variation is always present in some form after day 87.

The superhump appearing at day 87 is due to nodal precession at a beat period between the orbital and superhump periods. At day 87 the optical decline is reversed
in parallel with the hard (\textit{Swift}/BAT) X-rays. This would correspond to the growth of the radiative-warping triggered by the growth of the irradiating flux and probably 
due to increased accretion efficiency. The warp allows the surface of the secondary near the L$_1$ point to ``see'' the X-rays directly, which leads
to an enhanced mass-transfer rate \citep{Viallet08}. The impact of the enhanced  matter stream causes a deformation of the warp, thereby producing after day 103 
a modulation at $P_W$, that is closer to the orbital period. At day 114 the rapid drop in hard X-rays damps the effect of irradiation on the disc and the secondary, but enough flux is left
to continue to drive the warp. It is possible that the varying stream impact-radius modifies the warp structure.

Why does this happen at day 87?  We note that, at day 60, the NICER light curve (and hardness ratio) changes slope in a way typical to the cooling front beginning to propagate through the disc, and which usually signals the beginning of the end of an outburst. In this case the cooling front clearly fails to complete its job, and instead gets reflected back as a heating front, which happens when the inner disc radius is truncated and fixed \citep[see, eg.,][]{Dubus2001}. It seems that, by day 118, the inner disc radius reaches the ISCO, while the reflected front arrives at the outer disc
edge. The reflected-front propagation could therefore lead to disc expansion between days 87 - 118 \citep{Hameury20}.

\subsubsection{Irradiation}

From day 0 to day 87, the outburst is hard (for a ``soft'' X-ray transient), which means that, during the decay from maximum, a substantial ``corona'' is maintained.  It is most probably an ADAF (advection dominated accretion flow, e.g. \citealt{Esin1997}), i.e. the inner disc edge is (much) larger than the ISCO (innermost stable circular orbit, see \citealt{Dubus2001}). It is this ADAF that irradiates directly the outer disc regions with hard X-rays. The outer disc at some point becomes tilted and warped (see above), and if the irradiation is quasi--isotropic this would lead to a strong modulation at the precession period if the hard X-ray emitting region is high enough above the disc. This could happen if the steady jet (which has been present throughout the hard state, as evidenced by the radio flux in Figure \ref{fig:LC}) were a significant contributor to the X-ray flux as well, but this is still unclear \citep[see e.g.,][]{Shaw21}.

The level of tilting required can be crudely estimated if we simply take the far half of the disc (as observed at $\sim$phase 0.9) and tilt it towards us.  Requiring that this gives an optical increase of 0.5 mag can be achieved with a tilt of 23$^\circ$, relative to the measured binary inclination of $\sim$63$^\circ$.  We note that this ought to bring the disc close to producing dips in the X-ray flux at appropriate phases, and some have been seen \citep[e.g.,][]{Kajava19}, but this should be calculated with more realistic and detailed geometry \citep[as in e.g.,][]{Ogilvie01}.  

Maintaining the ADAF even near maximum is untypical, as one expects (at least from the model) the inner disc to then be at the ISCO. However. if a strong magnetic field is accumulating in the inner disc region, this could prevent it from moving in towards the BH.  There is strong evidence that the inner disc radius is moving in slowly, which comes from the detailed analysis of X-ray QPOs throughout the outburst by \citet{Stiele2020}.  They see the QPOs remaining at very low frequencies, between 0.1 and 0.5Hz, throughout most of the hard state, but they move rapidly to 4Hz at day 117, having doubled in frequency in just one day -- the truncated inner disc has moved in rapidly.   

Therefore during decay from maximum, the ADAF cools down and the field accumulates at the BH, creating what is called a MAD \citep[Magnetically Arrested Disc,][]{Tchekhovskoy11}, i.e. a configuration favouring launching jets by the Blandford--Znajek mechanism \citep{Blandford77}. At day 87, an outflow is created (a ``pre--jet-ejection'') forming a new source of hard X-rays, now high enough above the disc to irradiate the outer disc.  Such a source is referred to by \citet{deMarco21} as shocks within a ballistic jet, further above the disc. We note that \citet{Buisson19}'s fitting of the NuSTAR spectra taken on day 110 give a much greater height ($>$100 $R_g$ for the upper corona (``lamp-post'') compared to earlier spectra, and this could be related to the jet component, as also noted in the jet outflow of the Insight/HXMT spectral fits by \citet{You21}.  This outflow radiates anisotropically, which, combined with the disc warp, can now produce the observed modulation at the ``superhump'' period.  {\it This is what changes at day 87: the geometry of disc irradiation.} Between days 114 and 117 the outflow becomes an ejection \citep[seen by][]{Bright2020} which leaves the vicinity of the BH, leading to the ``collapse'' of the magnetic field, which allows the inner disc to reach the ISCO, and the transition to the soft state occurs from $\sim$day 118. 

The 87d timescale is also interesting given that it is comparable to the ``secondary maximum'' that has been seen in many LMXB XRTs \citep{Chen1993}. We believe that it is likely to be linked to these same processes involving both the inner and outer regions of the disc.

\section{Conclusions}

The spectacularly extensive and detailed AAVSO observations of J1820 provide the most complete optical outburst light-curve of any Black-Hole X-ray transient to date.  Combined with contemporaneous X-ray monitoring they offer a superb probe of the evolving geometry of both the inner and outer accretion disc regions.  Our key results and conclusions are:

\begin{itemize}
    
    \item these data show that the appearance of a large optical modulation on $P_{\rm sh}$ at day 87 is linked to the beginning of changes in the X-ray spectral properties of J1820;
    \item the large amplitude of this modulation ($\geq$0.6 mag) is too large to be explained via an area-variation effect in an eccentric, precessing disc.  This also calls into question this explanation for superhumps that have been seen in other LMXB XRTs;
    \item this modulation $\it cannot$ be due to an X-ray variation, as no X-ray modulation is present on any of the observed optical periodicities;
    \item the period of this optical modulation drifts on a timescale of $\sim$14d to ${\sim}P_W$, very close to $P_{\rm orb}$, and the variations in the light-curves show clearly that it cannot be explained as X-ray heating of the inner face of the donor.
    \item instead we interpret this effect as irradiation-driven warping of the outer disc, thereby creating sufficient disc area, tilted towards the observer to explain the modulation.
    \item this also requires a raised, hard X-ray emitting source that we associate with the outflow and base of the jet as it approaches the end of the hard state.
    \item much of our interpretation is speculative because detailed modelling of such behaviour is missing, largely because data of the quality and extent obtained for the J1820 outburst has simply not been available hitherto.  Accordingly, we hope that investigations into the structure and evolution of radiatively warped discs in X-ray binaries are worth a renewed effort.
\end{itemize}

 Future such campaigns should focus in particular on the 2--3 weeks around an X-ray state transition.  Whilst straightforward for ground-based monitoring facilities such as those offered by AAVSO and CBA, it can be difficult to anticipate such state changes so as to arrange for suitable X-ray coverage.  This will be particularly suited for future wide-field X-ray missions.

\section*{Data Availability}

The AAVSO optical data used throughout this paper are available from \url{https://aavso.org/aavso-international-database-aid}, and the radio data from the links in \citet{Bright2020}.
The X-ray data underlying this article are available from the following archives:
\begin{itemize}
    \item NICER --  \href{https://heasarc.gsfc.nasa.gov/docs/archive.html}{https://heasarc.gsfc.nasa.gov/docs/archive.html}
    \item Swift -- \href{https://www.swift.ac.uk/archive/}{https://www.swift.ac.uk/archive/}
    \item MAXI -- \href{http://maxi.riken.jp/top/index.html}{http://maxi.riken.jp/star\_data/J1820+071/J1820+071.html}

\end{itemize}

\section*{Acknowledgements}
We thank the anonymous referee whose comments and queries have led to significant improvements in the clarity and presentation of our results.
We acknowledge with thanks the variable star observations from the AAVSO International Database contributed by dedicated observers worldwide and used in this research. This work has also made use of data supplied by the UK Swift Science Data Centre at the University of Leicester.
The SALT observations were obtained under the SALT Large Science Programme on transients (2018-2-LSP-001; PI: DAHB) which is also supported by Poland under grant no. MNiSW DIR/WK/2016/07. DAHB and SBP acknowledge research support from the National Research Foundation.
JPL was supported in part by a grant from the French Space Agency CNES, and would like to thank Joe Smak for inspiring discussions on superhump behaviour.
JAP acknowledges support from STFC and a UGC-UKIERI Thematic Partnership, and is part supported by a University of Southampton Central VC Scholarship.


\bibliographystyle{mnras}
\bibliography{bibliography} 

\end{document}